\def\year{2019}\relax
\documentclass[letterpaper]{article} % DO NOT CHANGE THIS
\usepackage{aaai19}  % DO NOT CHANGE THIS
\usepackage{times}  % DO NOT CHANGE THIS
\usepackage{helvet} % DO NOT CHANGE THIS
\usepackage{courier}  % DO NOT CHANGE THIS
\usepackage[hyphens]{url}  % DO NOT CHANGE THIS
\usepackage[framemethod=TikZ]{mdframed}
\mdfdefinestyle{MyFrame}{%
    linecolor=black,
    outerlinewidth=1pt,
    roundcorner=10pt,
    innertopmargin=\baselineskip,
    innerbottommargin=\baselineskip,
    innerrightmargin=10pt,
    innerleftmargin=10pt,
    backgroundcolor=gray!20!white}
\usepackage{graphicx} % DO NOT CHANGE THIS
\usepackage{tabu, multirow, multicol,booktabs,ctable}
\usepackage{graphicx}  % DO NOT CHANGE THIS
\newcommand*\samethanks[1][\value{footnote}]{\footnotemark[#1]}
\usepackage[percent]{overpic}
\usepackage{amsmath}

\title{Uncovering Coordinated Networks on Social Media:\\ Methods and Case Studies}
\author{Diogo Pacheco,\thanks{Equal contributions.}\textsuperscript{\rm 1,2}
Pik-Mai Hui,\samethanks\textsuperscript{\rm 1}
Christopher Torres-Lugo,\samethanks\textsuperscript{\rm 1}\\
\Large \textbf{Bao Tran Truong,\textsuperscript{\rm 1} Alessandro Flammini,\textsuperscript{\rm 1} Filippo Menczer\textsuperscript{\rm 1}} \\ % All authors must be in the same font size and format. Use \Large and \textbf to achieve this result when breaking a line
\textsuperscript{\rm 1}Observatory on Social Media, Indiana University Bloomington, USA \\
\textsuperscript{\rm 2}Department of Computer Science, University of Exeter, UK 
}
 
\begin{document}

\maketitle

\begin{abstract}
Coordinated campaigns are used to influence and manipulate social media platforms and their users, a critical challenge to the free exchange of information online. Here we introduce a general, unsupervised network-based methodology to uncover groups of accounts that are likely coordinated. The proposed method constructs coordination networks based on arbitrary behavioral traces shared among accounts. We present five case studies of influence campaigns, four of which in the diverse contexts of U.S. elections, Hong Kong protests, the Syrian civil war, and cryptocurrency manipulation. In each of these cases, we detect networks of coordinated Twitter accounts by examining their identities, images, hashtag sequences, retweets, or temporal patterns. The proposed approach proves to be broadly applicable to uncover different kinds of coordination across information warfare scenarios.
\end{abstract}

\section{Introduction}

Online social media have revolutionized how people access news and information, and form opinions. By enabling exchanges that are unhindered by geographical barriers, and by lowering the cost of information production and consumption, social media have enormously broadened participation in civil and political discourse. Although this could potentially strengthen democratic processes, there is increasing evidence of malicious actors polluting the information ecosystem with disinformation and manipulation campaigns~\cite{Lazer-fake-news-2018,vosoughi2018spread,bessi2016social,Shao2018,ferrara2017disinformation,stella2018bots,deb2019perils,Bovet2019,Grinberg2019}.

While influence campaigns, misinformation, and propaganda have always existed~\cite{Jowett2018}, social media have created new vulnerabilities and abuse opportunities. Just as easily as like-minded users can connect in support of legitimate causes, so can groups with fringe, conspiratorial, or extremist beliefs reach critical mass and become impervious to expert or moderating views. Platform APIs and commoditized fake accounts make it simple to develop software to impersonate users and hide the identity of those who control these social bots --- whether they are fraudsters pushing spam, political operatives amplifying misleading narratives, or nation-states waging online warfare~\cite{ferrara2016rise}. Cognitive and social biases make us even more vulnerable to manipulation by social bots: limited attention facilitates the spread of unchecked claims, confirmation bias makes us disregard facts, group-think and echo chambers distort perceptions of norms, and the bandwagon effect makes us pay attention to bot-amplified memes~\cite{Weng2012,HillsProliferation18,Ciampaglia2018,Lazer-fake-news-2018,pennycook2019understanding}. 

Despite advances in countermeasures such as machine learning algorithms and human fact-checkers employed by social media platforms to detect misinformation and inauthentic accounts, malicious actors continue to effectively deceive the public, amplify misinformation, and drive polarization~\cite{barrett2019}.
We observe an arms race in which the sophistication of attacks evolves to evade detection. 

Most machine learning tools to combat online abuse target the detection of social bots, and mainly use methods that focus on individual accounts~\cite{davis2016botornot,varol2017online,Yang2019botometer,Yang2020botometer-lite,botometerv4-2020}. However, malicious groups may employ \textit{coordination} tactics that appear innocuous at the individual level, and whose suspicious behaviors can be detected only when observing networks of interactions among accounts. For instance, an account changing its handle might be normal, but a group of accounts switching their names in rotation is unlikely to be coincidental.

Here we propose an approach to reveal coordinated behaviors among multiple actors, regardless of their automated/organic nature or malicious/benign intent. The idea is to extract features from social media data to build a coordination network, where two accounts have a strong tie if they display unexpectedly similar behaviors. These similarities can stem from any metadata, such as content entities and profile features. Networks provide an efficient representation for sparse similarity matrices, and a natural way to detect significant clusters of coordinated accounts. 
% We demonstrate the effectiveness of the proposed approach on Twitter, but the method can in principle be applied to any social media platform where data is available. Since the method is completely unsupervised, no labeled training data is required. 
%
% After reviewing background literature and describing our methodology, we present five case studies by instantiating the approach to detect different types of coordination: (i)~handle changes, (ii)~image sharing, (iii)~sequential use of hashtags, (iv)~co-retweets, and (v)~synchronization. These examples illustrate the generality of our approach: we are able to detect coordinated campaigns based on what is presented as identity, shown in pictures, written in text, retweeted, or when these actions are taken.\footnote{Note to reviewers: when the paper is accepted, we will publish anonymized datasets in a public data repository to ensure reproducibility.} 
Our main contributions are:

\begin{itemize}
    \item We present a general approach to detect coordination, which can in principle be applied to any social media platform where data is available. Since the method is completely unsupervised, no labeled training data is required.
    
    \item Using Twitter data, we present five case studies by instantiating the approach to detect different types of coordination based on (i)~handle changes, (ii)~image sharing, (iii)~sequential use of hashtags, (iv)~co-retweets, and (v)~synchronization. 
    
    \item The case studies illustrate the generality and effectiveness of our approach: we are able to detect coordinated campaigns based on what is presented as identity, shown in pictures, written in text, retweeted, or when these actions are taken.
    
    \item We show that coordinated behavior does not necessarily imply automation. In the case studies, we detected a mix of likely bot and human accounts working together in malicious campaigns.
    
    \item Code and data are available at \url{github.com/IUNetSci/coordination-detection} to reproduce the present results and apply our methodology to other cases.

\end{itemize}

\section{Related Work}

Inauthentic coordination on social media can occur among social bots as well as human-controlled accounts. However, most research to date has focused on detecting social bots~\cite{ferrara2016rise}. 
%Early efforts in this area leveraged crowd-sourcing~\cite{wang2012social}.  However, automated approaches had to be developed to keep pace with the ever-increasing amount of social bots facilitated by their low cost of deployment. This led to the development of machine learning models, initially based on the supervised learning paradigm. These
Supervised machine learning models require labeled data describing how both humans and bots behave. 
%Given the lack of ground truth data, 
Researchers created datasets using automated honeypot methods~\cite{lee2011seven}, human annotation~\cite{varol2017online}, or likely botnets~\cite{echeverria2017discoveryA,echeverria2017discoveryB}. These datasets have proven useful in training supervised models for bot detection~\cite{davis2016botornot,varol2017online,Yang2019botometer}.

One downside of supervised detection methods is that by relying on features from a single account or tweet, they are not as effective at detecting coordinated accounts. This limitation has been explored in the context of detecting coordinated social bots~\cite{chen2018unsupervised,cresci2017paradigm,grimme2018changing}. 
%A coordinated social bot by itself might not seem suspicious, however, they become highly suspicious when considered in conjunction with the set of accounts they are coordinated with. 
The detection of coordinated accounts requires a shift toward the unsupervised learning paradigm. Initial applications focused on clustering or community detection algorithms in an attempt to identify similar features among pairs of accounts~\cite{ahmed2013generic,miller2014twitter}. Recent applications look at specific coordination dimensions, such as content or time~\cite{al2019deviance}. A method named \textit{Digital DNA} proposed to encode the tweet type or content as a string, which was then used to identify the longest common substring between accounts~\cite{cresci2016dna}. \textit{SynchroTrap}~\cite{cao2014uncovering} and \textit{Debot}~\cite{chavoshi2016debot} leverage temporal information to identify clusters of accounts that tweet in synchrony. 
Content-based methods proposed by \citeauthor{chen2018unsupervised}~(\citeyear{chen2018unsupervised}) and \citeauthor{giglietto2020}~(\citeyear{giglietto2020}) consider co-sharing of links on Twitter and Facebook, respectively. 
%An Ising model approach was proposed following the observation that bots exhibit heterophily in their interactions, in contrast to homophily among humans~\cite{mesnards2018detecting}.
Timestamp and content similarity were both used to identify coordinated accounts during the 2012 election in South Korea~\cite{keller2017manipulate,keller2019political}. 
%Unlike this method, our approach does not require ground-truth data, but assumes that a practitioner has a conjecture about suspicious behaviors.

While these approaches can work well, each is designed to consider only one of the many possible coordination dimensions. Furthermore, they are focused on coordination features that are likely observed among automated accounts; inauthentic coordination among human-controlled accounts is also an important challenge.
The unsupervised approach proposed here is more general in allowing multiple similarity criteria that can detect human coordination in addition to bots. 
As we will show, several of the aforementioned unsupervised methods can be considered as special cases of the methodology proposed here. 

%The quadratic complexity problem is mitigated by imposing sparse network representations for the bipartite relationships between accounts and traces, and consequently the derived feature and projection networks.

\section{Methods}

% original diagram can be found at https://drive.google.com/file/d/1bkfz3RIdoWMHbukTU7vKSbmX1WbZu4fi/view?usp=sharing
\begin{figure*}
    \centering
    \includegraphics[width=0.8\textwidth]{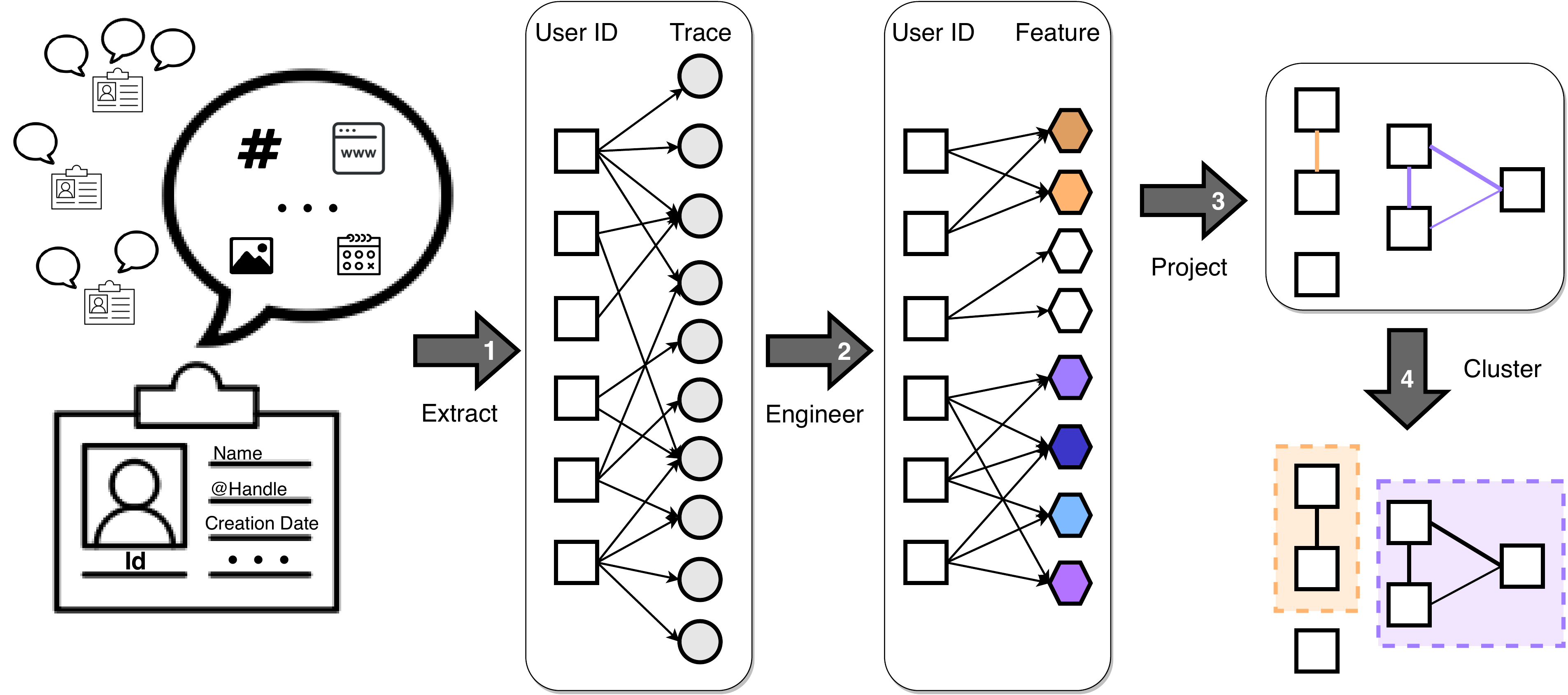}
    \caption{\textbf{Coordination Detection Approach}. On the left we see behavioral traces that can be extracted from social media profiles and messages. Four steps described in the text lead to the identification of suspicious clusters of accounts.} 
    \label{fig:detection_framework}
\end{figure*}

% \begin{figure}[t]
%     \centering
%     \includegraphics[width=.6\columnwidth]{images/example_tweet}
%     \includegraphics[width=.6\columnwidth]{images/example_user}
%     \includegraphics[width=.6\columnwidth]{images/element_types}
%     \includegraphics[width=.85\columnwidth]{images/framework}
%     \caption{\textbf{Coordination Detection Framework}. bla bla bla.} % Fil: call it metadata instead of element, use user instead of id
%     \label{fig:detection_framework}
% \end{figure}

The proposed approach to detect accounts acting in coordination on social media is illustrated in Fig.~\ref{fig:detection_framework}. It can be described by four phases:

\begin{enumerate}
    \item \textbf{Behavioral trace extraction:} The starting point of coordination detection should be a \emph{conjecture} about suspicious behavior. Assuming that authentic users are somewhat independent of each other, we consider a surprising lack of independence as evidence of coordination. The implementation of the approach is guided by a choice of \emph{traces} that capture such suspicious behavior. For example, if we conjecture that accounts are controlled by an entity with the goal of amplifying the exposure of a disinformation source, we could extract shared URLs as traces. 
    Coordination scenarios may be associated with a few broad categories of suspicious traces:
    
        \begin{enumerate}
        \item Content: if the coordination is based on the \textit{content} being shared, suspicious traces may include words, n-grams, hashtags, media, links, user mentions, etc. 
        
        \item Activity: coordination could be revealed by spatio-temporal patterns of \textit{activity}. Examples of traces that can reveal suspicious behaviors are timestamps, places, and geo-coordinates. 
        
        \item Identity: accounts could coordinate on the basis of personas or groups. Traces of \textit{identity} descriptors could be used to detect these kinds of coordination: name, handle, description, profile picture, homepage, account creation date, etc.
        
        \item Combination: the detection of coordination might require a \textit{combination} of multiple dimensions. For instance, instead of tracing only which hashtags were used or when accounts were active, as would be the case for a content- or activity-based suspicious trace, one can combine both these traces to have a temporal-content detection approach. The combined version is more restrictive and, therefore, can reduce the number of false positives.
    \end{enumerate}
    Once traces of interest are identified, we can build a network of accounts based on similar behavioral traces. 
    Preliminary data cleaning may be applied, filtering nodes with lack of \emph{support} --- low activity or few interactions with the chosen traces --- because of insufficient evidence to establish their coordination.
    For example, an account sharing few images will not allow a reliable calculation of image-based similarity. 
    %Similarly, one could not state with confidence that an account tweeting twice has been coordinating with anyone due to lack of data. 

    \item \textbf{Bipartite network construction:} 
    % Once traces of interest are identified, we can build a network of users based on similar behavioral traces. Preliminary data cleaning may be applied, for example filtering nodes with low activity --- we cannot state with confidence that an account tweeting twice has been coordinating with anyone due to lack of data. 
    The next step is to build a bipartite network connecting accounts and features extracted from their profiles and messages. In this phase, we may use the behavioral traces as features, or \emph{engineer new features} derived from the traces. For example, content analysis may yield features based on sentiment, stance, and narrative frames. Temporal features such as hour-of-day and day-of-week could be extrapolated from timestamp metadata. Features could be engineered by aggregating traces, for example by conflating locations into countries or images into color profiles. More complex features could be engineered by considering sets or sequences of traces. The \emph{bipartite network} may be \emph{weighted} based on the strength of association between an account and a feature --- sharing the same image many times is a stronger signal than sharing it just once. Weights may incorporate normalization such as IDF to account for popular features; it is not suspicious if many accounts mention the same celebrity. 
    % Let us define \emph{support} for an account as the degree or strength (weighted degree) of the corresponding node in the bipartite network, for example the number of tweets or images shared. Accounts with low support may be filtered at this stage because of insufficient evidence to establish their coordination --- an account sharing few images will not allow a reliable calculation of image-based similarity. 
    % optional steps, define a general rule here or explain in the case studies:
    % engineer traces?
    %     yes
    %     no
    % filtering low activity nodes:
    %     when?
    %     when not?
    % weights on the bipartite network:
    %     no weight?
    %     normalization, IDF?
    % filtering based on support (eg edge weight)?
    %     when?
    %     when not?
    
    \item \textbf{Projection onto account network:} 
    The bipartite network is projected onto a network where the account nodes are preserved, and edges are added between nodes based on some similarity measure over the features. The \emph{weight} of an edge in the resulting undirected \emph{coordination network} may be computed via simple co-occurrence, Jaccard coefficient, cosine similarity, or more sophisticated statistical metrics such as mutual information or $\chi^2$. 
    %The edges in the resulting undirected network are weighted by the similarity measure and reveal  interaction patterns among the accounts.
    In some cases, every edge in the coordination network is suspicious by construction. In other cases, edges may provide noisy signals about coordination among accounts, leading to false positives. For example, accounts sharing several of the same memes are not necessarily suspicious if those memes are very popular. In these cases, manual curation may be needed to \emph{filter out} low-weight edges in the coordination network to focus on the most suspicious interactions. One way to do this is to preserve edges with a top percentile of weights. The Discussion section presents edge weight distributions is some case studies, illustrating how aggressive filtering allows one to prioritize precision over recall, thus minimizing false positives. 
    %More advanced methods like multi-scaling backbone~\cite{serrano2009extracting} can also be used to filter edges. 
     
    \item \textbf{Cluster analysis:} The final step is to \emph{find groups} of accounts whose actions are likely coordinated on the account network. Network community detection algorithms that can be used for this purpose include connected components, $k$-core, $k$-cliques, modularity maximization, and label propagation, among others~\cite{fortunato2010community}. In the case studies presented here we use connected components because we only consider suspicious edges (by design or by filtering).

\end{enumerate}

In summary, the four phases of the proposed approach to detect coordination are translated into eight actionable steps: (i)~formulate a conjecture for suspicious behavior; (ii)~choose traces of such behavior, or (iii)~engineer features if necessary; (iv)~pre-filter the dataset based on support; choose (v)~a weight for the bipartite network and (vi)~a similarity measure as weight for the account coordination network; (vii)~filter out low-weight edges; and finally, (viii)~extract the coordinated groups.
Although the proposed method is unsupervised and therefore does not required labeled training data, we recommend a manual inspection of the suspicious clusters and their content. Such analysis will provide validation of the method and evidence of whether the coordinated groups are malicious and/or automated.

In the following sections we present five case studies, in which we implement the proposed approach to detect coordination through shared identities, images, hashtag sequences, co-retweets, and activity patterns. 

\section{Case Study 1: Account Handle Sharing}

% \begin{mdframed}[style=MyFrame]

% \begin{itemize}

% \item Conjecture: identities should not be shared, who are sharing them?

% \item Support: removing account with less than ten Botometer checks.

% \item Trace: screen name.

% \item Engineered trace: no.

% \item Bipartite weight: NA, the bipartite is unweighted.

% \item Projection weight: co-occurrence.

% \item Filtering edges: no.

% \item Clustering: connected components.

% \end{itemize}

% \end{mdframed}

%\subsection{Context}

On Twitter and some other social media platforms, although each user account has an immutable ID, many relationships are based on an account handle (called \texttt{screen\_name} on Twitter) that is changeable and in general reusable. An exception is that handles of suspended accounts are not reusable on Twitter. Users may have legitimate reasons for changing handles. However, the possibility of changing and reusing handles exposes users to abuse such as username squatting\footnote{\url{help.twitter.com/en/rules-and-policies/twitter-username-squatting}} and impersonation~\cite{Mariconti2017}. In a recent example, multiple Twitter handles associated with different personas were used by the same Twitter account to spread the name of the Ukraine whistleblower in the US presidential impeachment case.\footnote{\url{www.bloomberg.com/news/articles/2019-12-28/trump-names-ukraine-whistle-blower-in-a-retweet-he-later-deleted}} 

For a concrete example of how handle changes can be exploited, consider the following chronological events:
\begin{enumerate}
    \item \texttt{user\_1} (named \texttt{@super\_cat}) follows \texttt{user\_2} (named \texttt{@kittie}) who posts pictures of felines.
    
    \item \texttt{user\_3} (named \texttt{@super\_dog}) post pictures of canines.
    
    \item \label{item:tw}\texttt{user\_1} tweets mentioning \texttt{user\_2}: "I love \underline{\texttt{@kittie}}". A mention on Twitter creates a link to the mentioned account profile. Therefore, at time step 3, \texttt{user\_1}'s tweet is linked to \texttt{user\_2}'s profile page.
    
    \item \texttt{user\_2} renames its handle to \texttt{@tiger}.
    
    \item \texttt{user\_3} renames its handle to \texttt{@kittie}, reusing \texttt{user\_2}'s handle.
\end{enumerate}
Even though \texttt{user\_1}'s social network is unaltered regardless of the name change (\texttt{user\_1} still follows \texttt{user\_2}), name changes are not reflected in previous posts, so anyone who clicks on the link at step \ref{item:tw} will be redirected to \texttt{user\_3}'s profile instead of to \texttt{user\_2} as originally intended by \texttt{user\_1}. This type of user squatting, in coordination with multiple accounts, can be used to promote entities, run ``follow-back'' campaigns, infiltrate communities, or even promote polarization~\cite{Mariconti2017}. Since social media posts are often indexed by search engines, these manipulations can be used to promote content beyond social media boundaries.

To detect this kind of coordination on Twitter, we applied our approach using identity traces, namely Twitter handles.
We started from a log of requests to \url{Botometer.org}, a social bot detection service of the Indiana University Observatory on Social Media~\cite{Yang2019botometer}. 
Each log record consists of a timestamp, the Twitter \texttt{user\_id} and handle, and the bot score. We focus on users with at least ten entries (queries) such that multiple handle changes could be observed. This yielded 54 million records with
%from February 2017 to April 2019, containing 1.8 million unique accounts (each with a distinct \texttt{user\_id}) and 
1.9 million handles. For further details see Table~\ref{tab:case1}.

\subsection{Coordination Detection}

We create a bipartite network of suspicious handles and accounts. We consider a handle suspicious if it is shared by at least two accounts, and an account suspicious when it has taken at least one suspicious handle. Therefore no edges are filtered. 
One could be more restrictive, for example by considering
%a handle suspicious if it is shared by more than two accounts or
an account suspicious if it has taken more than one suspicious handle. 
%However, since coincidental handle collision is extremely rare, we use the minimal threshold to detect more and larger coordinated groups.  
To detect the suspicious clusters we project the network, connecting accounts based on the number of times they shared a handle. This is equivalent to using co-occurrence, the simplest similarity measure.
Each connected component in the resulting network identifies a cluster of coordinated accounts as well as the set of handles they shared. 
%Community detection algorithms can be applied to find sub-groups within the each component.
Table~\ref{tab:case1} summarizes the method decisions.

\begin{table}
    \small
    \centering
    \caption{Case Study 1 summary}
    \label{tab:case1}
    \begin{tabular}{r|l}
    \toprule
    Conjecture & Identities should not be shared\\
    Support filter & Accounts with $<10$ records\\
    Trace & Screen name\\
    Eng. trace & No\\
    Bipartite weight & NA, the bipartite is unweighted\\ 
    Proj. weight & Co-occurrence\\
    Edge filter & No\\
    Clustering & Connected components\\
    Data source & Botometer~\cite{Yang2019botometer}\\
    Data period &  Feb 2017--Apr 2019\\
    No. accounts & 1,545,892\\
    \bottomrule
    \end{tabular}
\end{table}

\subsection{Analysis}
% results

% \begin{figure*}
%     \centering
%     \includegraphics[trim = 5mm 4mm 5mm 8mm, clip,width=.35\textwidth]{images/suspicious_edges_labeled-user_proj-v2_and_v3-all_components}
%     \includegraphics[trim = 5mm 60mm 5mm 60mm, clip,width=.55\textwidth]{images/suspicious_edges_labeled-user_proj-v2_and_v3-giant_components}
% \caption{\textbf{Shared \texttt{handle} network shows suspicious coordinated groups}. Nodes represent 2,757 Twitter accounts, in 253 connected components, and the weight of their connection is the number of times two accounts switched names. The giant component is highlighted on the right, the 722 accounts colored according to their communities and edges showing the shared handles.}
%     \label{fig:screen_name_network}
% \end{figure*}

\begin{figure*}[t]
% https://drive.google.com/file/d/1tLHjXzmP4i8Ql7rh3ipGryy4lXq-EdT3/view?usp=sharing
    \centering
    \includegraphics[width=.8\textwidth]{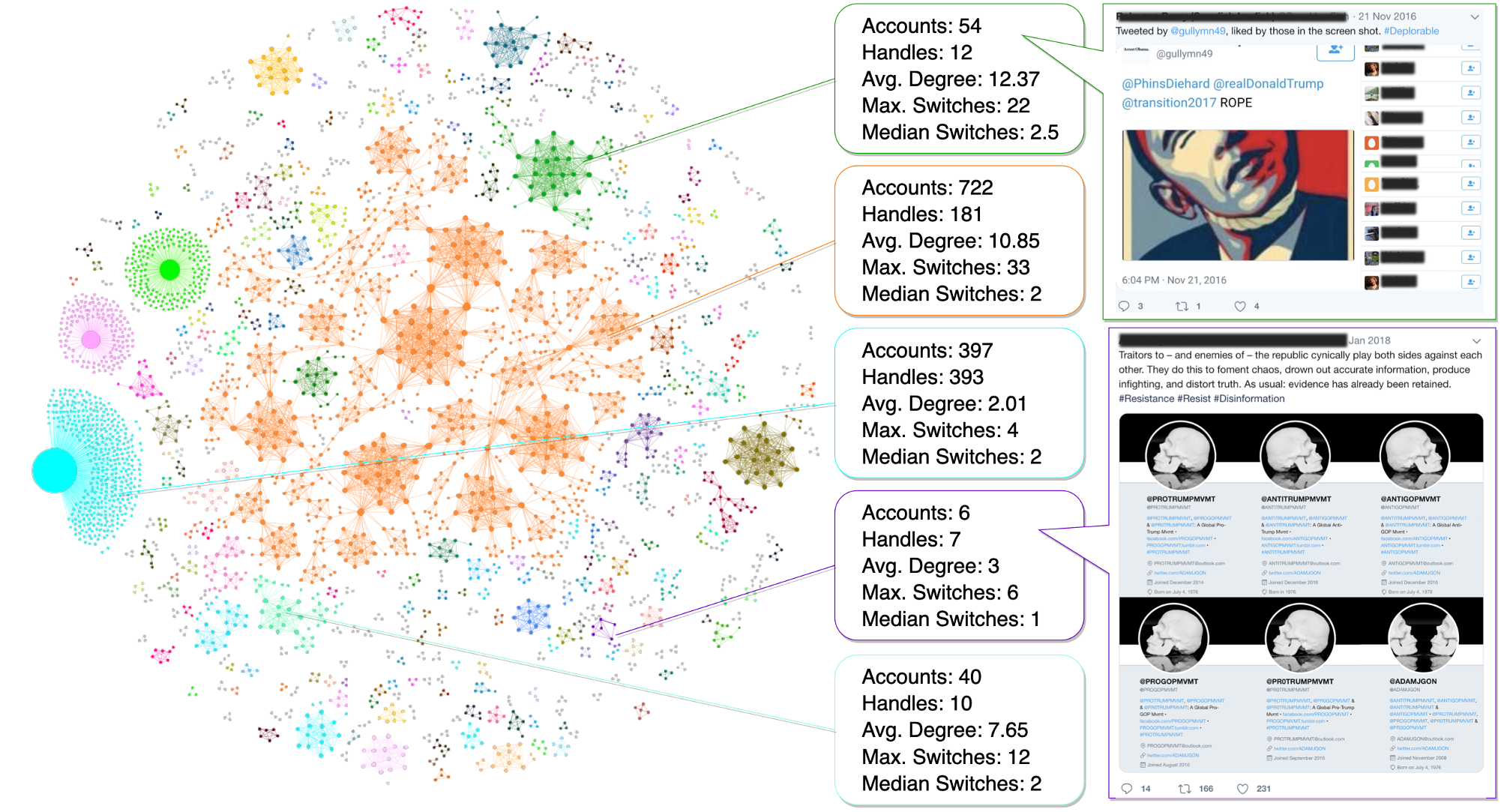}
\caption{\textbf{Handle sharing network.} A node represents a Twitter account and its size is proportional to the number of accounts with which it shares handles. The weight of an edge is the number of unique handles shared by two accounts. Suspicious coordinated groups are identified by different colors. We illustrate the characteristics of a few coordinated groups, namely the number of accounts, number of shared handles, average number of accounts with which handles are shared, and the maximum and median number of times that a handle is switched among accounts. The number of switches is a lower-bound estimate on the basis of our data sample. We also show tweets by independent parties who uncovered the malicious activity of a couple of the coordinated groups, discussed in the main text.}
    \label{fig:screen_name_network}
\end{figure*}

Fig.~\ref{fig:screen_name_network} shows the handle sharing network. It is a weighted, undirected network with 7{,}879 nodes (Twitter accounts). We can classify the components into three classes:

\begin{enumerate}

    \item \textbf{Star-like components} capture the major accounts (hub nodes) practicing name squatting and/or hijacking. To confirm this, we analyzed the temporal sequence of handle switches involving star-like components. Typically, a handle switches from an account (presumably the victim) to the hub, and later (presumably after some form of ransom is paid) it switches back from the hub to the original account. These kinds of reciprocal switches occur 12 times more often in stars than any other components.
    % An edge-weight of two means two accounts switched their names twice (a \textit{double-change} edge); these two names can be the same---temporary change---or not. Malicious successful attempts of exchanging name are characterized by a \textit{single-name-double-change} edge. Double-change edges are 12 times more likely to be found in the 3 star-like components than in the other ones. Moreover, among double-change edges, 100\% are single-named in the stars, but only 54\% in the non-stars. 
    
    \item \textbf{The giant component} includes 722 accounts sharing 181 names (orange group in the center of Fig.~\ref{fig:screen_name_network}). Using the Louvain community detection algorithm~\cite{blondel2008fast}, we further divide the giant component into 13 sub-groups. We suspect they represent temporal clusters corresponding to different coordinated campaigns by the same group. This investigation is left for future study.  
    
    \item \textbf{Other components} can represent different cases requiring further investigation, as discussed next.

\end{enumerate}

Fig.~\ref{fig:screen_name_network} illustrates a couple of stories about malicious behaviors corresponding to two of the coordinated handle sharing groups, which was uncovered by others. 
In June 2015, the handle \texttt{@GullyMN49} was reported in the news due to an offensive tweet against President Obama.\footnote{minnesota.cbslocal.com/2015/06/03/obama-tweeter-says-posts-cost-him-his-job-2/} More than one year later, the same handle was still posting similar content.
%\footnote{\url{twitter.com/SwedJewFish/status/800946662386503680}} 
In March 2017, we observed 23 different accounts taking the handle in a 5-day interval. We conjecture that this may have been an attempt to keep the persona created back in 2015 alive and evade suspension by Twitter following reports of abuse to the platform. Currently, the \texttt{@GullyMN49} handle is banned but 21 of the 23 accounts are still active. 
    
The second example in Fig.~\ref{fig:screen_name_network} shows a cluster of six accounts sharing seven handles. They have all been suspended since. Interestingly, the cluster was sharing handles that appeared to belong to conflicting political groups, e.g., \texttt{@ProTrumpMvmt} and \texttt{@AntiTrumpMvmt}. Some of the suspicious accounts kept changing sides over time. Further investigation revealed that these accounts were heavily active; they created the appearance of political fundraising campaigns in an attempt to take money from both sides.

% \begin{figure}
%     \centering
%     \begin{overpic}[trim = 5mm 74mm 5mm 70mm, clip,width=.95\columnwidth]{images/suspicious_user_proj-v2_and_v3-all_network}
%       \put (58,12) {\includegraphics[width=.18\columnwidth]{images/suspicious_user_proj-v2_and_v3-edge_weight}}
%       \put (78,12) {\includegraphics[width=.18\columnwidth]{images/suspicious_user_proj-v2_and_v3-component_size}}
%     \end{overpic}
%     \caption{\textbf{Shared \texttt{handle} network shows suspicious coordinated groups}. Nodes represent 2,757 Twitter accounts, in 253 connected components, and the weight of their connection is the number of times two accounts switched names. In the bottom, the CCDF of edge-weight (left) and component-size (right).}
%     \label{fig:screen_name_network}
% \end{figure}

% \begin{figure}
%     \centering
%     \includegraphics[trim = 5mm 25mm 4mm 36mm, clip,width=.95\columnwidth]{images/suspicious_edges_labeled-user_proj-v2_and_v3-comp88-trump}
%     % \vspace{2cm}
%     \includegraphics[trim = 5mm 18mm 4mm 6mm, clip,width=.95\columnwidth]{images/suspicious_edges_labeled-user_proj-v2_and_v3-comp88-trump-timeline} 
% \caption{\textbf{Changing sides}. In the top, a suspicious components: nodes are accounts and edges represent the sequence of handles shared between two accounts. In the bottom, nodes are represented as timelines showing which handles they used (y-axis).}
%     \label{fig:component88-trump}
% \end{figure}

% Temporal Analysis

\section{Case Study 2: Image Coordination}

% \subsection{Context}

Images constitute a large portion of the content on social media.
A group of accounts posting many of the same or similar images may reveal suspicious coordinated behavior. In this case study, we identify such groups on Twitter in the context of the 2019 Hong Kong protest movement by leveraging media images as content traces. We used the BotSlayer tool~\cite{huibotslayer} to collect tweets matching a couple dozen hashtags related to the protest in six languages, and subsequently downloaded all images and thumbnails in the collected tweets. 
We focus on 31{,}772 tweets that contain one or more images, and remove all retweets to avoid trivial replications of the same images.
%The remaining 2,945 users generated 31,772 tweets with images.
More on the data source can be found in Table~\ref{tab:case2}.

\begin{figure*}[t]
\centering
\includegraphics[width=0.8\textwidth]{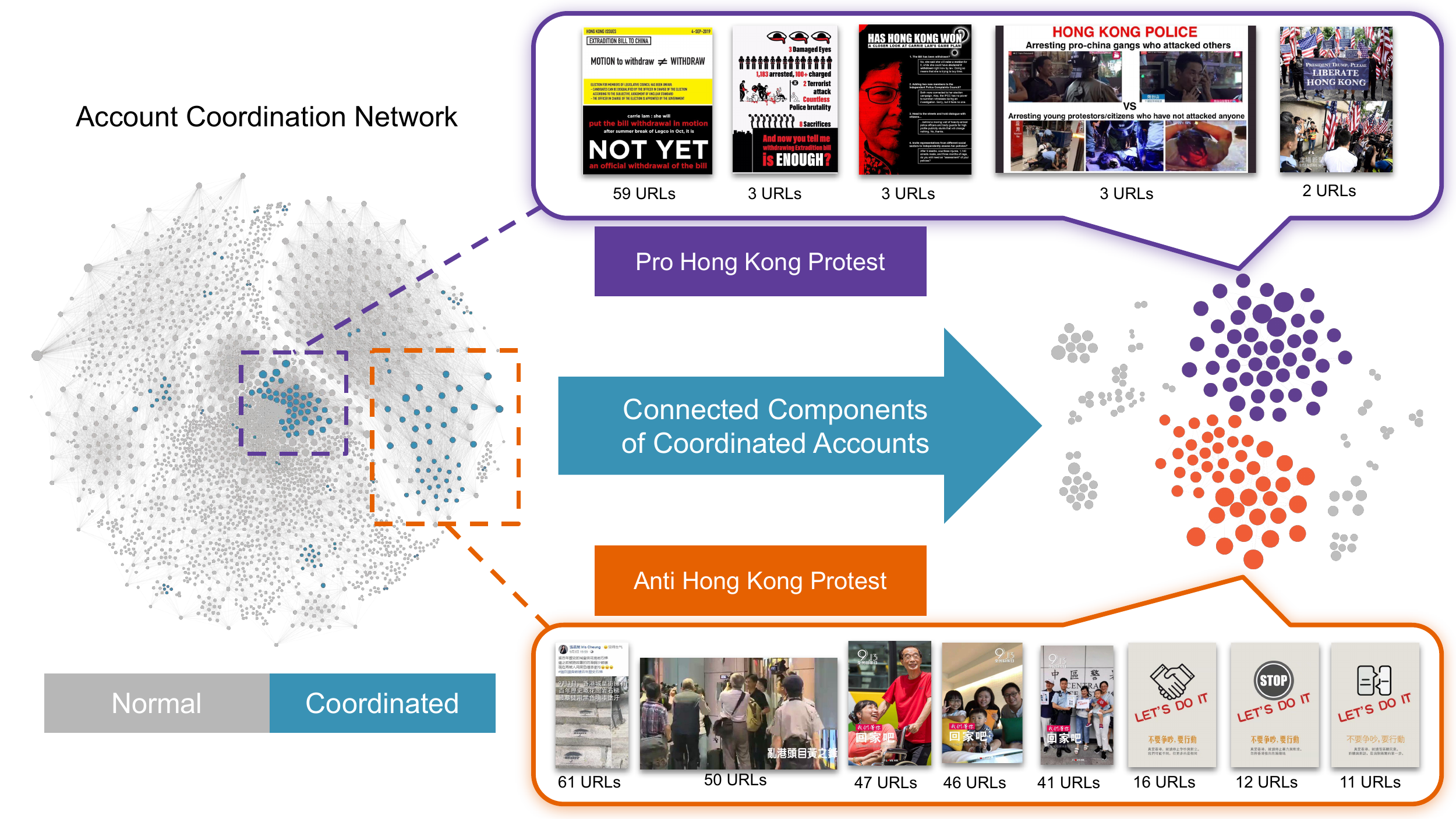}
\caption{\textbf{Account coordination network about Hong Kong protest on Twitter.} Nodes represent accounts, whose sizes are proportional to their degrees. On the left-hand side, accounts are colored blue if they are likely coordinated, otherwise gray. On the right-hand side we focus on the connected components corresponding to the likely coordinated groups. The three largest components are colored according to the content of their images --- one pro- and two anti-protest clusters, in purple and orange respectively. We show some exemplar images shared by these groups, along with the corresponding numbers of distinct URLs.}
\label{fig:hkp_case_study}
\end{figure*}

\subsection{Coordination Detection}

Every time an image is posted, it is assigned a different URL.
Therefore detecting identical or similar images is not as simple as comparing URLs; it is necessary to analyze the actual image content.
We represent each image by its RGB color histogram, binning each channel into 128 intervals and resulting in a 384-dimensional vector.
The binned histograms allow for matching variants: images with the same vector are either identical or similar, and correspond to the same feature.
While enlarging the bins would give more matches of variants, we want to ensure the space is sparse enough to retain high match precision. 
% added as description of effect of bin size --PM

We exclude accounts who tweeted less than five images to reduce noise from insufficient support. 
One could tune precision and recall by adjusting this support threshold. 
We set the threshold to maximize precision while maintaining reasonable recall. The sensitivity of precision to the support threshold parameter is analyzed in the Discussion section.

We then construct an unweighted bipartite network of accounts and image features by linking accounts with the vectors of their shared images.
%The network edges are weighted by their corresponding tweet frequencies. 
We project the bipartite network to obtain a weighted account coordination network, with edge weights computed by 
%any similarity measure calculated using the bipartite network weights; in this case study we adopt 
the Jaccard coefficient.
%, a simple and efficient similarity measure for selecting suspicious edges. 
%
We consider accounts that are highly similar in sharing the same images as coordinated.
To this end, we retain the edges with the largest 1\% of the weights (see Fig.~\ref{fig:weightdist}). 
%to focus on reliable coordination relations. % (~0.6667).
Excluding the singletons (accounts with no evidence of coordination), we rank the connected components of the network by size.
Table~\ref{tab:case2} summarizes the method decisions in this case.

\begin{table}
    \small
    \centering
    \caption{Case Study 2 summary}
    \label{tab:case2}
    \begin{tabular}{r|l}
    \toprule
    Conjecture & Unlike to upload duplicated images\\
    Support filter & Accounts with $<5$ tweets w/image\\
    Trace & Raw images\\
    Eng. trace & RBG intervals (128 bins on each ch.)\\
    Bipartite weight & NA, the bipartite is unweighted\\ 
    Proj. weight & Jaccard similarity\\
    Edge filter & Keep top 1\% weights\\
    Clustering & Connected components\\
    Data source & BotSlayer~\cite{huibotslayer}\\
    Data period & Aug--Sep 2018\\
    No. accounts & 2,945\\
    \bottomrule
    \end{tabular}
\end{table}

%Next, let us focus on the largest components. % in an effort to reveal the most suspicious groups.

\subsection{Analysis}

Fig.~\ref{fig:hkp_case_study} shows the account coordination network. 
We identify three suspicious clusters involving 315 accounts, posting pro- or anti-protest images.
The anti-protest group shares images with Chinese text, targeting Chinese-speaking audiences, while the pro-protest group shares images with English text.

We observe that some of the shared image features correspond to the exact same image, others are slight variants. For example, the 59 image URLs corresponding to the same feature in the pro-protest cluster include slight variations with different brightness and cropping. The same is true for 61 corresponding anti-protest images.

Although this method identifies coordination of accounts, it does not characterize the coordination as malicious or benign, nor as automated or organic. 
In fact, many of these coordinated accounts behave like humans (see Discussion). 
These groups are identified because their constituent accounts have circulated the same sets of pictorial content significantly more often than the rest of the population.

\section{Case Study 3: Hashtag Sequences}

\begin{figure}
\centering
\includegraphics[width=\columnwidth]{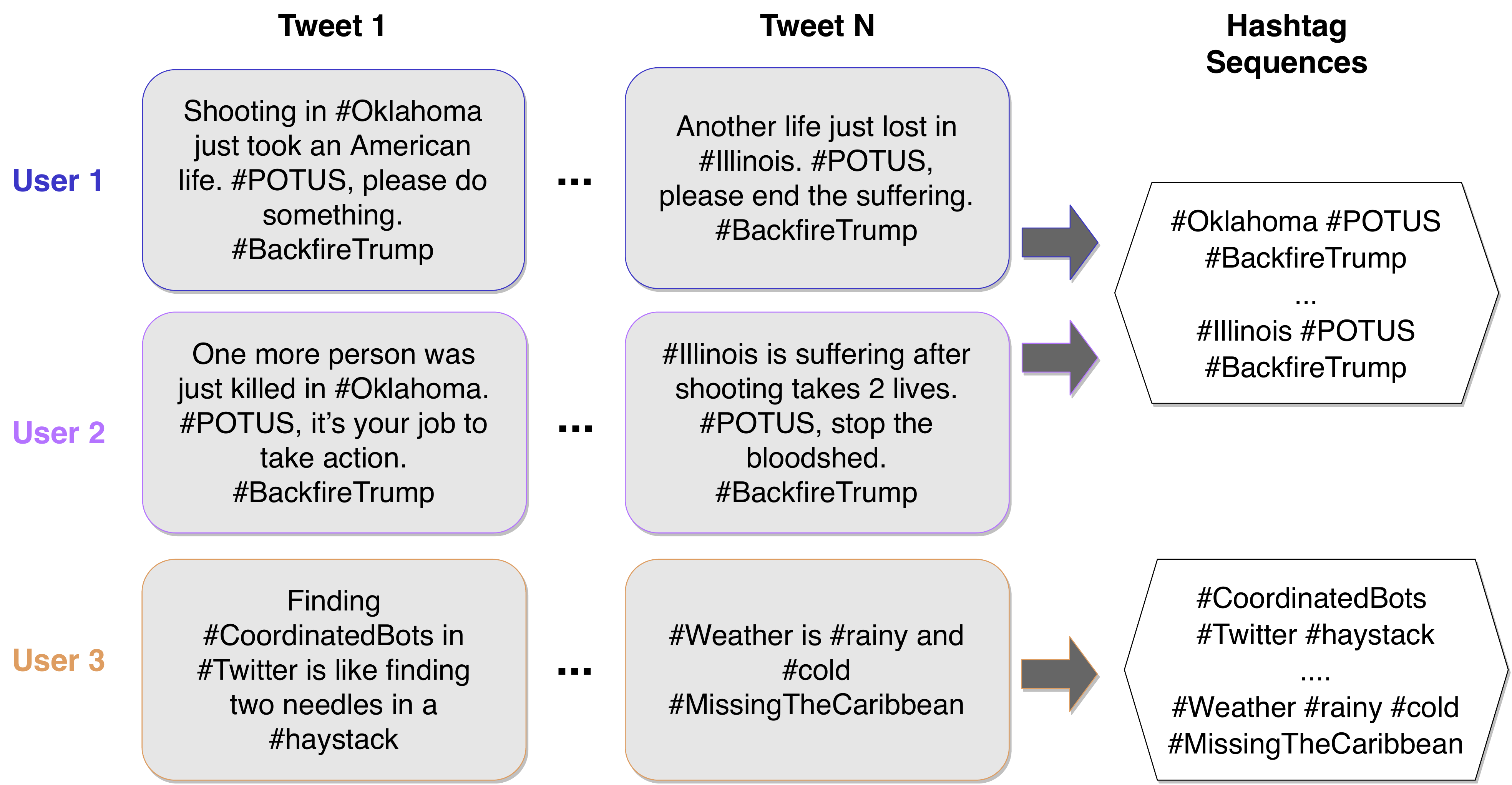}
\caption{\textbf{Hashtag sequence features.} Hashtags and their positions are extracted from tweet metadata. Accounts tweeting the same sequence of hashtags are easily identified.}
\label{fig:hashtag-seq}
\end{figure}

\begin{figure}
\centering
\includegraphics[width=0.95\columnwidth]{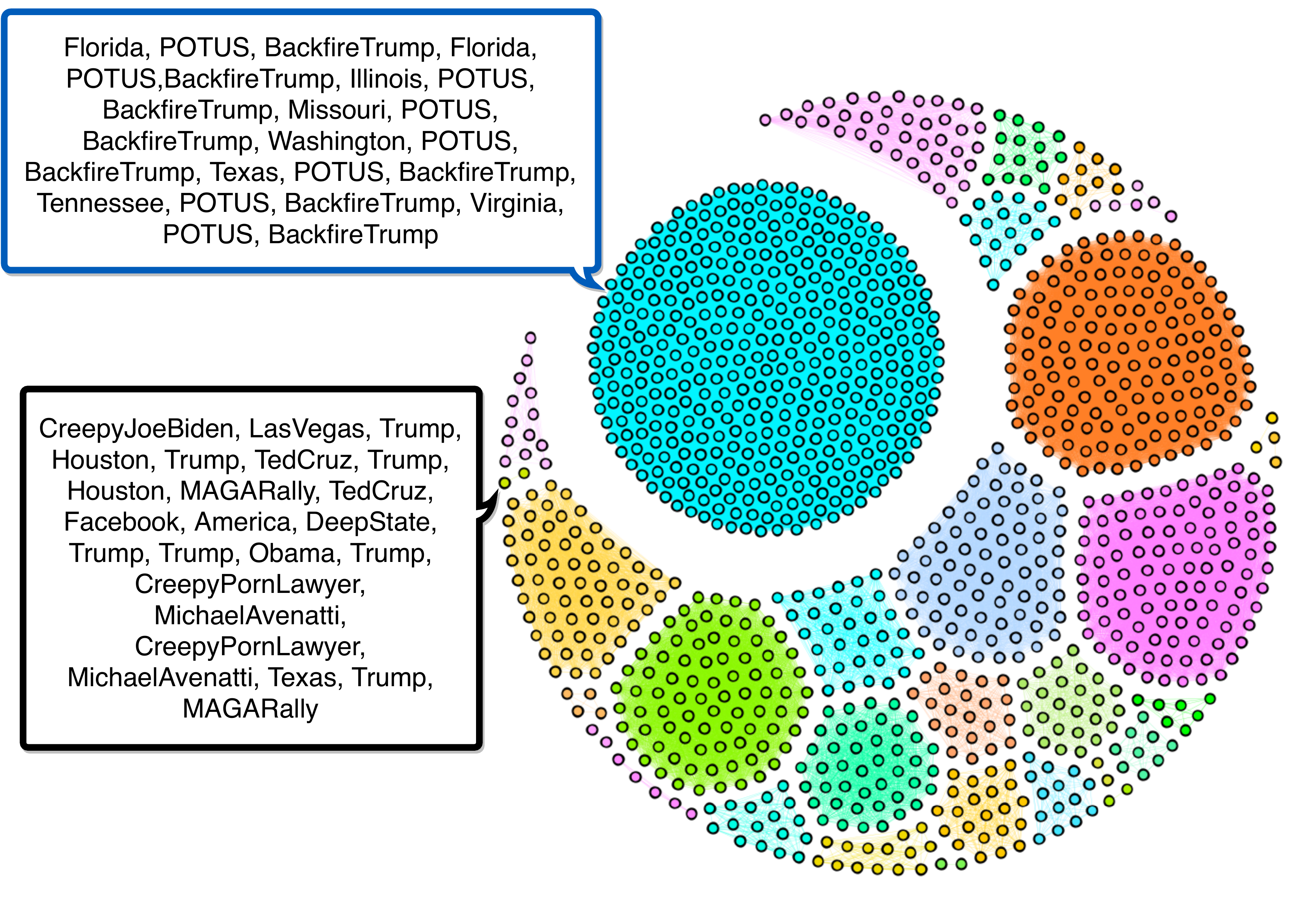}
\caption{\textbf{Hashtag coordination network.} Accounts are represented as nodes, with edges connecting accounts that tweeted the same sequences of hashtags. There are 32 connected components, identified by different colors. The hashtag sequences shared by two of the coordinated groups (the smallest and largest) are shown. This network is based on tweets from October 22, 2018.}
\label{fig:hashtag-network}
\end{figure}
% Link to edit Figure 7 in draw.io: https://drive.google.com/file/d/1W69KdxLcEWy6Szyu8ER75Kmaq6iw648r/view?usp=sharing

% \subsection{Context}

A key element of a disinformation campaign is an ample audience to influence. To spread beyond one's followers, a malicious actor can use hashtags to target other users who are interested in a topic and may search for related tweets. 

If a set of automated accounts were to publish messages using identical text, it would look suspicious and would be easily detected by a platform's anti-spam measures.
To minimize the chances of detection, it is easy to imagine a malicious user leveraging a language model (e.g., GPT-2\footnote{\url{openai.com/blog/better-language-models/}}) to paraphrase their messages. Detection could become harder due to apps that publish paraphrased text on behalf of a user. An example of this behavior is exhibited by the ``Backfire Trump'' Twitter app, which tweeted to President Trump whenever there was a fatality resulting from gun violence. 

However, we conjecture that even paraphrased text is likely to include the same hashtags based on the targets of a coordinated campaign. Therefore, in this case study we explore how to identify coordinated accounts that post highly similar sequences of hashtags across multiple tweets.

%\paragraph{Dataset} 
We evaluated this approach on a dataset of original tweets (no retweets) collected around the 2018 U.S. midterm election. More on the data source can be found in Table~\ref{tab:case3}. 
%
%We evaluated this approach on 
%the dataset generated for the Bot Electioneering Volume project \cite{yang2019bot}. The dataset consist 
%a dataset of tweets containing election-related hashtags, collected using Twitter's filtering API between October and December 2018, around the U.S. midterm election. 
Prior to applying our framework, we split the dataset into daily intervals to detect when pairs of accounts become coordinated. 

\subsection{Coordination Detection} %Build the Network}

% We create a bipartite network of users and their sequences of hashtags. A set of accounts is considered suspicious if they are tweeting the highly similar or the same sequence of different hashtags. To find whether two accounts are tweeting in this fashion, we order their tweets chronologically, and extract all the hashtags that they used in the order that they appeared on the tweet. We filter out users that do not meet either of these conditions: users with less than 5 tweets or users that had tweeted less than 5 tweets. 

A data preprocessing step filters out accounts with few tweets and hashtags. The thresholds depend on the time period under evaluation. In this case we use a minimum of five tweets and five unique hashtags over a period of 24 hours to ensure sufficient support for possible coordination. 
More stringent filtering could be applied to decrease the probability of two accounts producing similar sequences by chance.

In this case we engineer features that combine content (hashtags) and activity (timestamps) traces. In particular, we use \emph{ordered sequences} of hashtags for each user (Fig.~\ref{fig:hashtag-seq}). 
The bipartite network consists of accounts in one layer and hashtag sequences in the other. 
In the projection phase, we draw an edge between two accounts with identical hashtag sequences. These edges are unweighted and we do not apply any filtering, based on the assumption that two independent users are unlikely to post identical sequences of five or more hashtags on the same day. We also considered a fuzzy method to match accounts with slightly different sequences and found similar results.

%The duplicate detection method will fail to . It might also miss some coordinated accounts due to missing data. 
%Therefore we also consider a second approach, which uses a locality-sensitive hashing method such as MinHash~\cite{broder1997resemblance}. 
%This approach allows us the flexibility to detect similar users.  
%To construct the features, we use hashtag n-gram strings.\footnote{\url{github.com/ekzhu/datasketch}} The length of the n-grams can be tuned depending on the expected diversity of the hashtags used by suspicious accounts. 
%We project the MinHash bipartite network by computing the Jaccard similarity between sets of hashtag n-grams.
%We explored different thresholds to filter edges with low Jaccard similarity, and the results were robust to the threshold value.

%\paragraph{Find Clusters}
%A threshold of 0.8 on the similarity in the MinHash network yields results similar to using the exact-duplicate method; we present results based on the latter.  
We identify suspicious groups of accounts by removing singleton nodes and then extracting the connected components of the network. Large components are more suspicious, as it is less likely that many accounts post the same hashtag sequences by chance.
Table~\ref{tab:case3} summarizes the method decisions.

\begin{table}
    \small
    \centering
    \caption{Case Study 3 summary}
    \label{tab:case3}
    \begin{tabular}{r|l}
    \toprule
    Conjecture & Similar large sequence of hashtags\\
    Support filter & At least 5 tweets, 5 hashtags per day\\
    Trace & Hashtags in a tweet\\
    Eng. trace & Ordered sequence of hashtags in a day\\
    Bipartite weight & NA, the bipartite is unweighted\\ 
    Proj. weight & Co-occurrence\\
    Edge filter & No\\
    Clustering & Connected components\\
    Data source & BEV~\cite{yang2019bot}\\
    Data period &  Oct--Dec 2018\\
    No. accounts & 59,389,305\\
    \bottomrule
    \end{tabular}
\end{table}

\subsection{Analysis}

We identified 617 daily instances of coordination carried out by 1{,}809 unique accounts. 
Fig.~\ref{fig:hashtag-network} illustrates 32 suspicious groups identified on a single day. The largest component consists of 404 nodes that sent a series of tweets through the ``Backfire Trump'' Twitter application, advocating for stricter gun control laws.  This application no longer works. 
Some of the claims in these tweets are inconsistent with reports by the non-profit Gun Violence Archive. 
The smallest groups consist of just pairs of accounts.  
One of these pairs tweeted a link to a now-defunct page promoting bonuses for an online casino.
Another pair of accounts promoted a link to a list of candidates for elected office that had been endorsed by the Humane Society Legislative Fund.
% We observe that many of the coordinated accounts hijack hashtags to amplify their reach or use Twitter apps that tweet on behalf of an account. The latter is the case of the largest coordination group. 
One could of course use longer time windows and potentially reveal larger coordinated networks. For example, the Backfire Trump cluster in Fig.~\ref{fig:hashtag-network} is part of a larger network of 1{,}175 accounts.

%Since coordination can occur over multiple groups of accounts and these groups can evolve over time, it may be desirable to merge the daily networks to reveal more complex types of coordination among different groups. One could also use different time resolutions to build the network, possibly affecting the number of matches and false positives. 
%; shorter intervals enable more matches but also increase the noise, whereas longer intervals will yield fewer matches and fewer false positives.

% As a baseline, we compare our approach to using only Jaccard similarity over the set of hashtags tweeted by a user in a single day. In order have a sensible network size, we used graph intersection of the union of a sliding window of seven days over our dataset.

\section{Case Study 4: Co-Retweets}

Amplification of information sources is perhaps the most common form of manipulation. 
On Twitter, a group of accounts retweeting the same tweets or the same set of accounts may signal coordinated behavior. Therefore we focus on retweets in this case study.

%Dataset
We apply the proposed approach to detect coordinated accounts that amplify narratives related to the White Helmets, a volunteer organization that was targeted by disinformation campaigns during the civil war in Syria.\footnote{\url{www.theguardian.com/world/2017/dec/18/syria-white-helmets-conspiracy-theories}} Recent reports identify Russian sources behind these campaigns~\cite{wilson2020cross}. The data was collected from Twitter using English and Arabic keywords.  More details about the data can be found in Table~\ref{tab:case4}.
%The anonymized dataset in this case study is provided by the DARPA SocialSim project; it includes over 800 thousand retweets by approximately 42 thousand accounts, collected between April 2018 and March 2019.

\subsection{Coordination Detection}

%We focus on active accounts with at least ten retweets.
We construct the bipartite network between retweeting accounts and retweeted messages, excluding self-retweets and accounts having less than ten retweets.
This network is weighted using TF-IDF to discount the contributions of popular tweets.
Each account is therefore represented as a TF-IDF vector of retweeted tweet IDs.
The projected co-retweet network is then weighted by the cosine similarity between the account vectors.
Finally, to focus on evidence of potential coordination, we keep only the most suspicious 0.5\% of the edges (see Fig.~\ref{fig:weightdist}). These parameters can be tuned to trade off between precision and recall; the effect of the thresholds on the precision is analyzed in the Discussion section.
%Fig.~\ref{fig:precision}.
Table~\ref{tab:case4} summarizes the method decisions.

\begin{table}
    \small
    \centering
    \caption{Case Study 4 summary}
    \label{tab:case4}
    \begin{tabular}{r|l}
    \toprule
    Conjecture & High overlapping of retweets\\
    Support filter & Accounts with $<10$ retweets\\
    Trace & Retweeted tweet ID\\
    Eng. trace & No\\
    Bipartite weight & TF-IDF\\ 
    Proj. weight & Cosine similarity\\
    Edge filter & Keep top 0.5\% weights\\
    Clustering & Connected components\\
    Data source & DARPA SocialSim \\
    Data period &  Apr 2018--Mar 2019\\
    No. accounts & 11,669\\
    \bottomrule
    \end{tabular}
\end{table}

\subsection{Analysis}

\begin{figure}[t]
\centering
\includegraphics[width=\columnwidth]{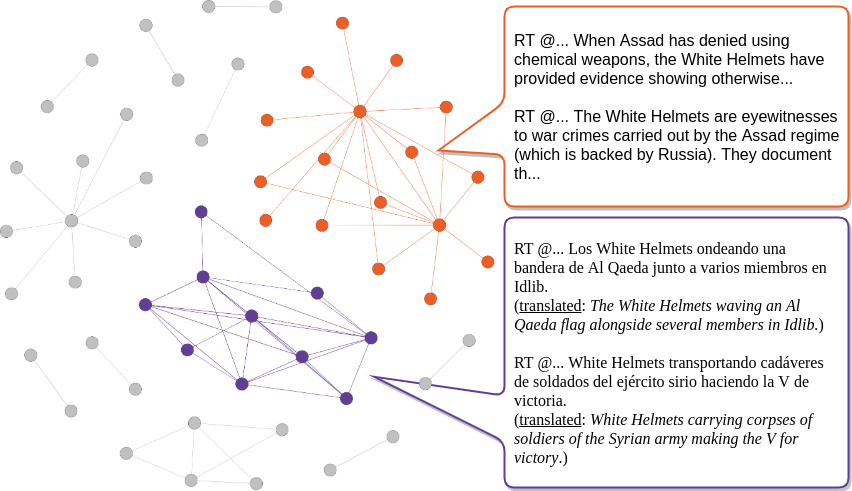}
\caption{\textbf{Co-retweet network.} Two groups are highlighted with exemplar retweets. Singletons are omitted.}
\label{fig:coretweet}
\end{figure}

Fig.~\ref{fig:coretweet} shows the co-retweet network, and highlights two groups of coordinated accounts.
Accounts in the orange and purple clusters retweet pro- and anti-White Helmets messages, respectively.
The example tweets shown in the figure are no longer publicly available.

\section{Case Study 5: Synchronized Actions}

\begin{figure*}
\centering
\includegraphics[width=1.5\columnwidth]{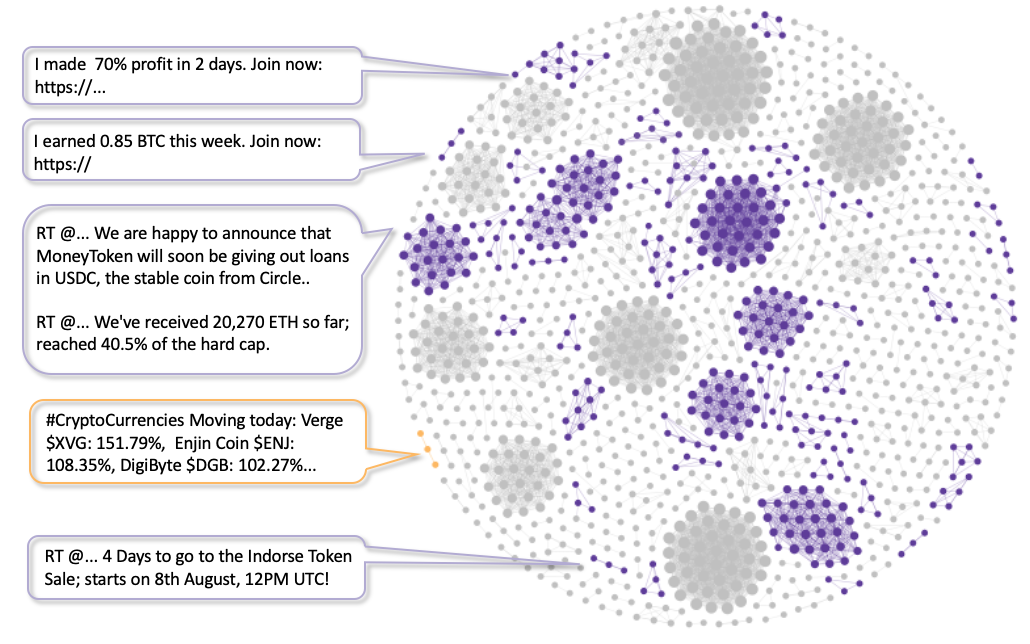}
\caption{\textbf{Time coordination network.} Nodes (accounts) are connected if they post or retweet within the same 30-minute periods. Singletons are omitted. Accounts in the purple clusters and the small yellow cluster at 8 o'clock are highly suspicious of running pump \& dump schemes. A few tweet excerpts are shown; these tweets are no longer publicly available.}
\label{fig:cotime}
\end{figure*}

``Pump \& dump" is a shady scheme where the price of a stock is inflated by simulating a surge in buyer interest through false statements (pump) to sell the cheaply purchased stock at a higher price (dump). Investors are vulnerable to this kind of manipulation because they want to act quickly when acquiring stocks that seem to promise high future profits. By exposing investors to information seemingly from different sources in a short period of time, fraudsters create a false sense of urgency that prompts victims to act.

Social media provides fertile grounds for this type of scam~\cite{mirtaheri2019identifying}. We investigate the effectiveness of our approach in detecting coordinated cryptocurrency pump \& dump campaigns on Twitter. The data was collected using keywords and cashtags (e.g., \$BTC) associated with 25 vulnerable cryptocoins as query terms. We consider both original tweets and retweets because they all add to the stream of information considered by potential buyers. More details on the dataset are found in Table~\ref{tab:case5}. 

\subsection{Coordination Detection}

We hypothesize that coordinated pump \& dump campaigns use software to have multiple accounts post pump messages in close temporal proximity. Tweet timestamps are therefore used as the behavioral trace of the accounts. 
The shorter the time interval in which two tweets are posted, the less likely they are to be coincidental. However, short time intervals result in significantly fewer matches and increased computation time. On the other hand, longer (e.g., daily) intervals produce many false positive matches. To balance between these concerns, we use 30-minute time intervals.

Intuitively, it is likely that any two users would post one or two tweets that fall within any time interval; however, the same is not true for a set of more tweets. To focus on accounts with sufficient support for coordination, we only keep those that post at least eight messages. This specific support threshold value is chosen to minimize false positive matches, as shown in the Discussion section.
%Fig.~\ref{fig:precision}. 

The tweets are then binned based on the time interval in which they are posted. These time features are used to construct the bipartite network of accounts and tweet times. Edges are weighted using TF-IDF. Similar to the previous case, the projected account coordination network is weighted by the cosine similarity between the TF-IDF vectors. Upon manual inspection, we found that many of the tweets being shared in this network are not related to cryptocurrencies, while only a small percentage of edges are about this topic. These edges also have high similarity and yield a strong signal of coordination. Thus, we only preserve the 0.5\% of edges with largest cosine similarity (see Fig.~\ref{fig:weightdist}). 
Table~\ref{tab:case5} summarizes the method decisions.

\begin{table}
    \small
    \centering
    \caption{Case Study 5 summary}
    \label{tab:case5}
    \begin{tabular}{r|l}
    \toprule
    Conjecture & Synchronous activities\\
    Support filter & Accounts with $<8$ tweets\\
    Trace & Tweet timestamp\\
    Eng. trace & 30-minute time intervals\\
    Bipartite weight & TF-IDF\\ 
    Proj. weight & Cosine similarity\\
    Edge filter & Keep top 0.5\% weights\\
    Clustering & Connected components\\
    Data source & DARPA SocialSim\\
    Data period &  Jan 2017--Jan 2019\\
    No. accounts & 887,239\\
    \bottomrule
    \end{tabular}
\end{table}

\subsection{Analysis}

Fig.~\ref{fig:cotime} shows the synchronized action network. The connected components in the network are qualitatively analyzed to evaluate precision. The purple subgraphs flag clusters of coordinated accounts where suspicious pump \& dump schemes are observed. We find different instances of this scheme for many cryptocurrencies. The excerpts included in Fig.~\ref{fig:cotime} are from tweets pushing the Indorse Token and Bitcoin, respectively. These tweets allegedly state that the accounts have access to business intelligence and hint at the potential rise in coin price. 

Changes in stock markets, especially those focusing on short-term trading such as cryptocurrencies, are hard to capture due to market volatility. Furthermore, it is difficult to attribute shifts in price to a single cause, such as pump \& dump-related Twitter activities. This makes it difficult to quantitatively validate our results. However, in the week of Dec 15--21, 2017 there were daily uptrends for the coins Verge (XVG), Enjin (ENJ), and DigiByte (DGB). On each day, the prices spiked after large volumes of synchronized tweets commenting on their moving prices. These trends preceded the record price for these coins to date, which was on Dec 23, 2017 for XVG, and Jan 7, 2018 for both ENJ and DGB. The  cluster of high-volume accounts pumping these three coins is highlighted in yellow in Fig.~\ref{fig:cotime}. 

Inspection of the dense clusters shown in gray in Fig.~\ref{fig:cotime} reveals they are composed of spam accounts or coordinated advertisement. Although not examples of pump \& dump schemes, they do correctly reflect coordinated manipulation.

\section{Discussion}

The five case studies presented in this paper are merely illustrations of how our proposed methodology can be implemented to find coordination. The approach can in principle be applied to other social media platforms besides Twitter.
For instance, the image coordination method can be applied on Instagram, and coordination among Facebook pages can be discovered via the content they share.

\begin{figure}[t]
    \centering
    \includegraphics[trim = 3mm 5mm 3mm 5mm, clip,width=.9\columnwidth]{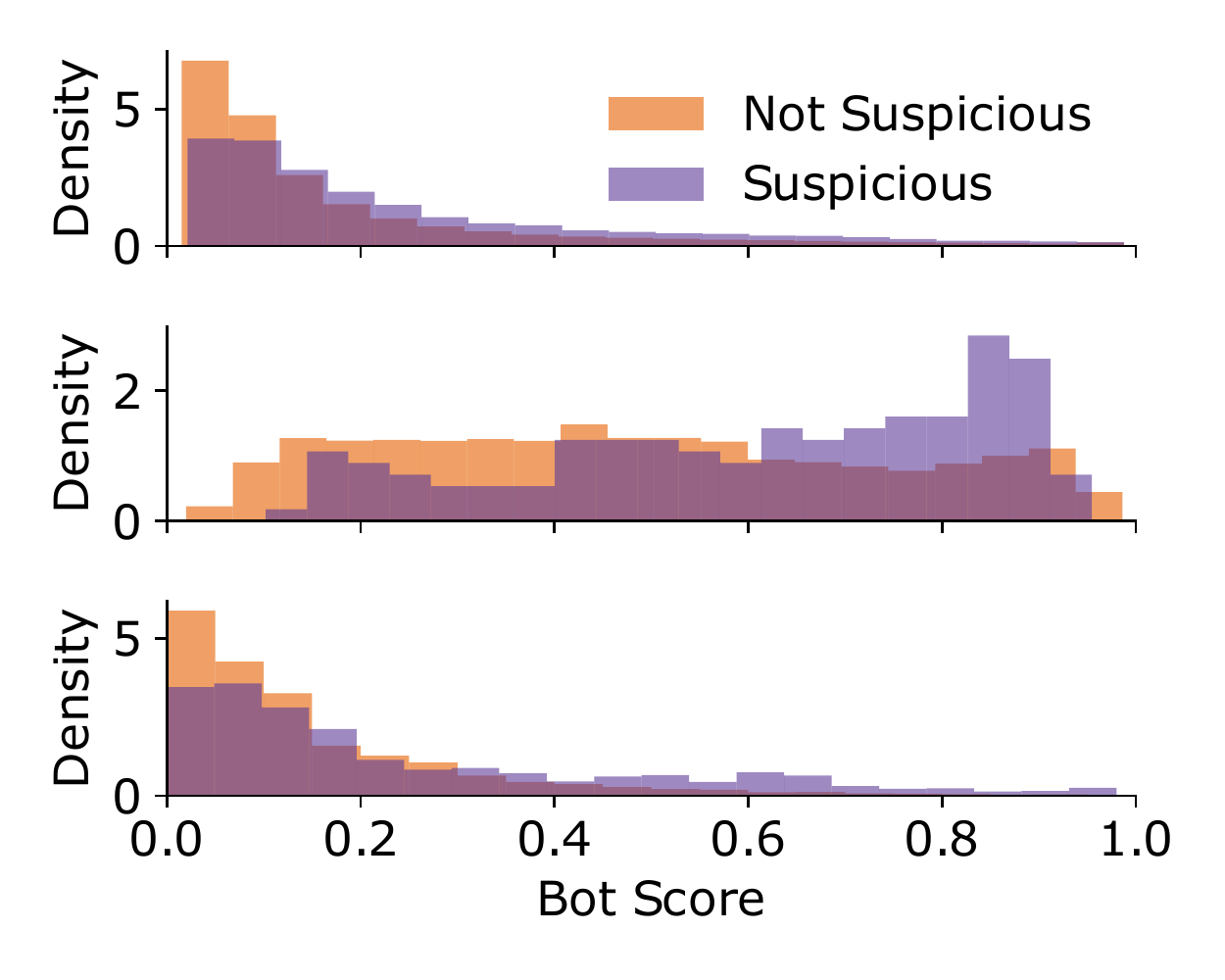}
    \caption{\textbf{Bot scores of suspicious and non-suspicious accounts.} Histograms of bot scores for the suspicious accounts identified by our methodology vs.~other accounts. Top, center, and bottom panels represent account handle sharing (Case Study 1), image coordination (Case Study 2), and hashtag sequence (Case Study 3), respectively.
    Bot scores for Case Study 1 are obtained from version 3 of Botometer~\cite{Yang2019botometer}, collected between May 2018 and April 2019. For the other two cases, the bot scores are obtained from BotometerLite~\cite{Yang2020botometer-lite}. 
    The datasets may include multiple scores for the same account. 
    }
    \label{fig:botscore_distribution}
\end{figure}

Several of the unsupervised methods discussed in the Related Work section, just like the five applications of our method presented here, focus on different types of coordination. These methods are therefore not directly comparable.  A key contribution of this paper is to provide a flexible and general methodology to describe these different approaches in a unified scheme. For example, Debot~\cite{chavoshi2016debot} can be described as a special case of our approach based on a sophisticated temporal hashing scheme preserving dynamic time warping distance~\cite{keogh2005exact}, while SynchroTrap~\cite{cao2014uncovering} exploits synchronization information by matching actions within time windows. The methods by \citeauthor{giglietto2020}~(\citeyear{giglietto2020}) and \citeauthor{chen2018unsupervised}~(\citeyear{chen2018unsupervised}) are special cases using similarity based on shared links. The method by \citeauthor{ahmed2013generic}~(\citeyear{ahmed2013generic}) uses a contingency table of accounts by features equivalent to our bipartite network. Finally, we explored the use of similar text within short time windows to detect coordinated networks of websites pretending to be independent news sources~\cite{pacheco2020whitehelmets}.

% \begin{figure}

% \begin{subfigure}[h]{0.4\linewidth}
% \includegraphics[width=.4\columnwidth]{images/suspicious_user_vs_normal-hist-cases-1_2_3.pdf}
% \caption{Bot score distribution}
% \end{subfigure}

% \hfill

% \begin{subfigure}[h]{0.4\linewidth}
% \includegraphics[width=\linewidth]{images/qqplot.pdf}
% \caption{Q-Q Plot}
% \end{subfigure}%

% \caption{\textbf{Bot scores of suspicious and non-suspicious accounts.} Histograms of the bot score for the suspicious populations (orange) identified by our methodology and for the non-suspicious accounts (purple). Top, center, and bottom panels represent \textit{account handle sharing}, \textit{image coordination}, and \textit{hashtag sequence} case studies, respectively.
% Bot scores obtained from the current version of the bot detection tool, collected since May 2018. The dataset may include multiple scores for the same account. 
% }
% \end{figure}

\begin{figure}[t]
    \centering
    \includegraphics[trim = 10mm 7mm 10mm 7mm, clip,width=.7\columnwidth]{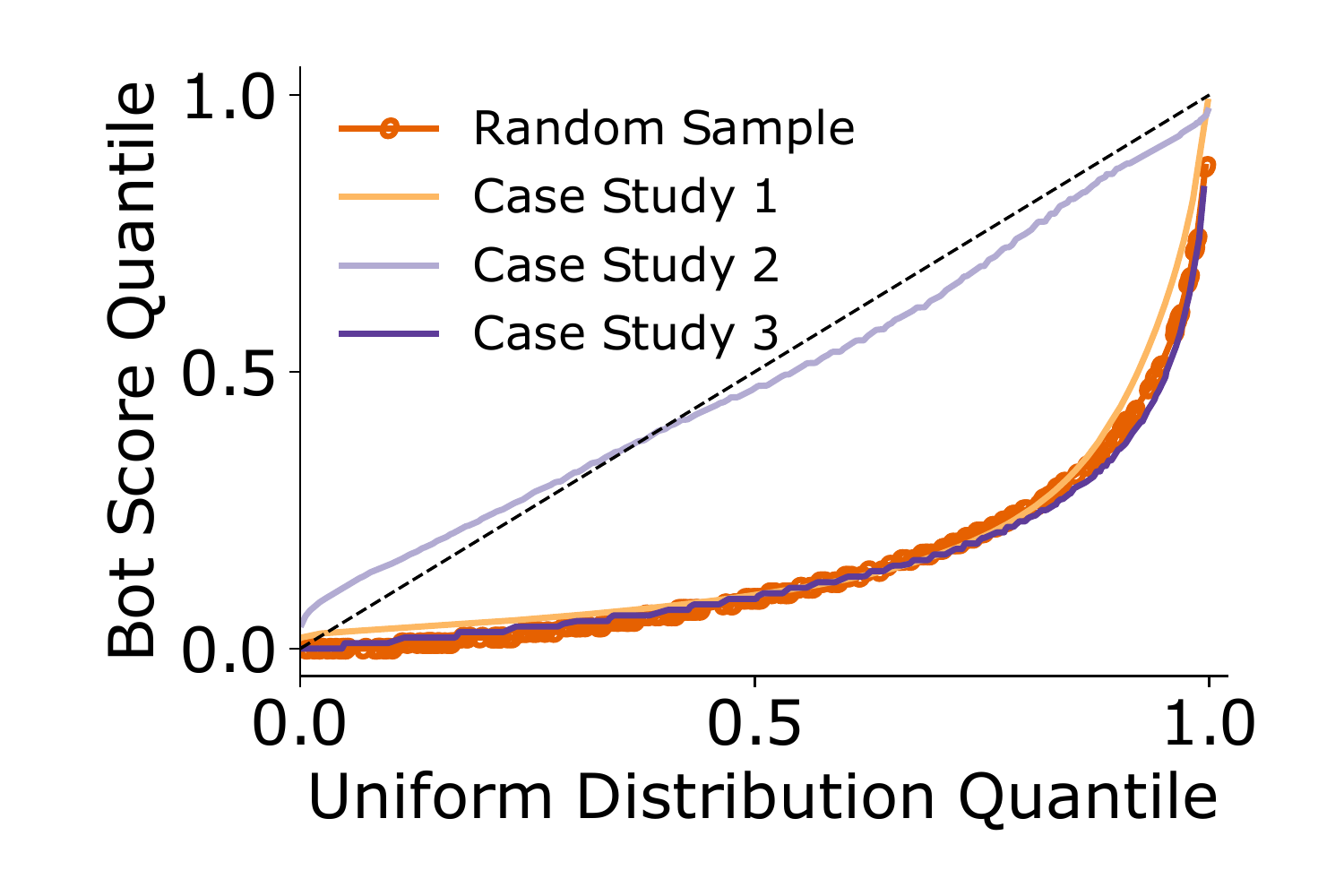}
    \caption{\textbf{Bot score distributions.} Q-Q plots comparing the distributions of bot scores in three case studies against that obtained from a 24-hour, 1\% random sample of tweets. The sources of the bot scores are explained in Fig.~\ref{fig:botscore_distribution}. All distributions are heavily skewed towards lower bot score values (i.e., more humans than bots), except Case Study 2 in which bot scores are higher, with a near-uniform distribution.}
    \label{fig:qqplot}
\end{figure}

Our approach aims to identify coordination between accounts, but it does not characterize the intent or authenticity of the coordination, nor does it allow to discover the underlying mechanisms. 
%A practitioner is assumed to start from a conjecture about specific suspicious behaviors and choose traces and features accordingly.
An example of malicious intent was highlighted in recent news reports about a coordinated network of teenagers posting false narratives about the election.\footnote{\url{www.washingtonpost.com/politics/turning-point-teens-disinformation-trump/2020/09/15/c84091ae-f20a-11ea-b796-2dd09962649c_story.html}}
However, it is important to keep in mind that coordinated campaigns may be carried out by authentic users with benign intent.
For instance, social movement participants use hashtags in a coordinated fashion to raise awareness of their cause. 

Fig.~\ref{fig:botscore_distribution} shows the distributions of bot scores in case studies 1--3. (We are unable to analyze bot scores in cases 4--5 due to anonymization in the datasets.) We observe that while coordinated accounts are more likely to have high bot scores, many coordinated accounts have low (human-like) scores --- the majority in two of the three cases. Therefore, detecting social bots is not sufficient to detect coordinated campaigns. 

\begin{figure}[t]
    \centering
    \includegraphics[trim = 4mm 5mm 4mm 1mm, clip, width=.75\columnwidth]{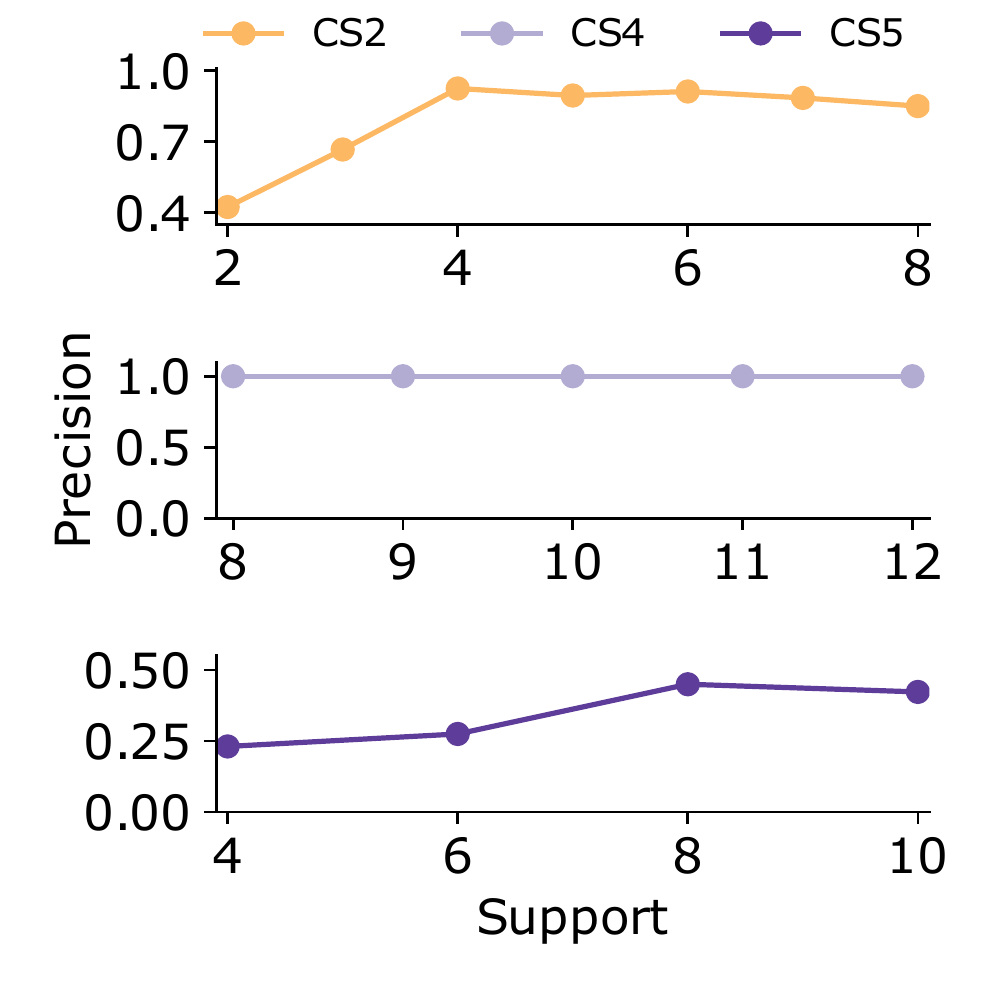}
    \caption{\textbf{Precision vs support.} In three of the case studies, we manually annotated accounts to calculate precision (fraction of accounts in suspicious clusters that are actually coordinated). Precision is plotted against support, namely, number of images (Case Study 2), number of retweets (Case Study 4), and number of tweets (Case Study 5).}
    \label{fig:precision}
\end{figure}

% dataset biases/representativeness
Although the case studies presented here are based on data from diverse sources, they were not designed to inflate the effectiveness of the proposed method, nor to focus on malicious accounts. Fig.~\ref{fig:qqplot} shows that the sets of accounts analyzed in case studies 1 and 3 have bot score distributions consistent with those obtained from a random sample of tweets. We note this is not a random sample of accounts --- it is biased by account activity. Case Study 2 is the exception; we conjecture that bots were used to post high volumes of images during the Hong Kong protest.

\begin{figure}
    \centering
    \includegraphics[width=.95\columnwidth]{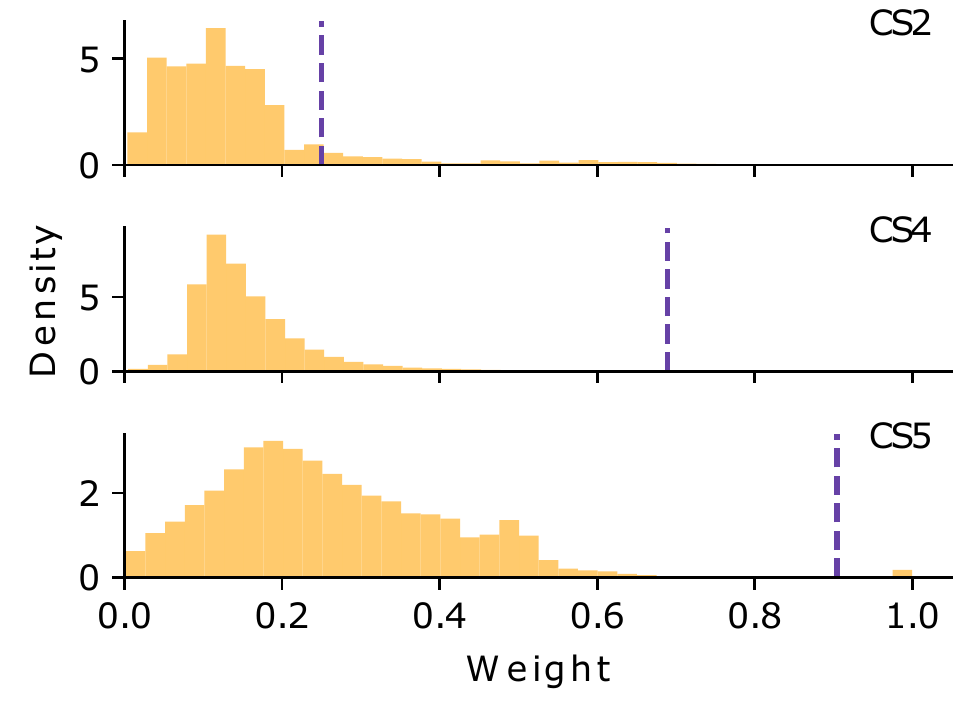}
    \caption{\textbf{Weight distributions in coordination networks for three case studies.} Dashed lines represent edge filters: we retain the edges with top 1\% of weights in Case 2 and top 0.5\% in Case 4 and 5.}
    \label{fig:weightdist}
\end{figure}

While our methodology is very general, each implementation involves design decisions and related parameter settings. 
%For example, in our case studies, likely coordinated groups are identified by simply considering connected components in the account coordination network. This may suffice if the network is sparse, whereas additional filtering (e.g., $k$-core or clique discovery) may help obtain better results in denser networks.
%Discuss about thresholds
These parameters can be tuned to trade off between false positive and false negative errors. In this paper we focus on minimizing false positives --- organic collective behaviors that may appear as coordinated. For instance, false positives could result from identical messages generated by social media share buttons on news websites; content similarity alone does not constitute evidence of coordination.  One way to avert false positives is to engineer features that are suspicious by construction, as in case studies 1 and 3. Another way is to filter accounts based on support thresholds, filter edges in the coordination network based on similarity, and filter clusters based on characteristics such as density or size. Fig.~\ref{fig:precision} illustrates how the support threshold affects precision in case studies 2, 4, and 5, based on manual annotations. Case Study 2 shows that support can be selected to maximize precision. In Case Study 4, precision is high irrespective of support because all accounts co-retweeting a high number of tweets are suspicious in that context. In Case Study 5, on the other hand, precision is low because the dataset contains a large portion of tweets unrelated to cryptocurrencies --- even though they too are coordinated. 
Fig.~\ref{fig:weightdist} illustrates the choices of edge filtering thresholds in the same case studies.  
%shortly after a tsunami hit Japan in Nov 2016, people came together to show their support by tweeting \texttt{\#PrayForJapan} and related hashtags.  

More rigorous methods could be investigated to exclude spurious links that can be attributed to chance. One approach we plan to explore in future work is to design null models for the observed behaviors, which in turn would enable a formal statistical test to identify meaningful coordination links. For example, one could apply Monte Carlo shuffling of the bipartite network before projection to calculate the $p$-values associated with each similarity link.

With the exception of the hashtag sequence feature that combines content and activity traces, our case studies explore single behaviors in their distinct contexts. Considering multiple dimensions could yield larger groups if the different types of coordination are related. On the other hand, independent types of coordination could be revealed by separate clusters. To illustrate this, combining co-retweets and shared URLs in Case Study 4 yields separate clusters, suggesting distinct and independent coordination campaigns. More work can be done in considering multiple dimensions of coordination in specific scenarios. This presents the challenge of representing interactions through multiplex networks, and/or combining different similarity measures.

\section{Conclusion}

In this paper we proposed a network approach to identify coordinated accounts on social media. We presented five case studies demonstrating that our approach can be applied to detect multiple types of coordination on Twitter. 

Unlike supervised methods that evaluate the features of individual accounts to estimate the likelihood that an account belongs to some class, say a bot or troll, our approach aims to detect coordinated behaviors at the group level. Therefore, the proposal is intended to complement rather than replace individual-level approaches to counter social media manipulation. Our method can also be leveraged to identify and characterize abusive accounts, which in turn can be used to train supervised learning algorithms.

The proposed approach provides a unified way of tackling the detection of coordinated campaigns on social media. As such, it may help advance research in this area by highlighting the similarities and differences between approaches.

We hope that this work will shed light on new techniques that social media companies may use to combat malicious actors, and also empower the general public to become more aware of the threats of modern information ecosystems.

BotSlayer~\cite{huibotslayer} lets users track narratives that are potentially amplified by coordinated campaigns. We plan to incorporate the framework presented in this paper into the BotSlayer system. We believe that the framework's flexibility, combined with the user-friendliness of BotSlayer, will enable a broader community of users to join our efforts in countering disinformation on social media.

\small

\paragraph{Acknowledgments.}

We thank Kaicheng Yang for helpful discussion.
We gratefully acknowledge support from the Knight Foundation, Craig Newmark Philanthropies, DARPA (contract W911NF-17-C-0094), and NSF (NRT award 1735095).

\bibliographystyle{aaai}
\bibliography{ms}

\begin{thebibliography}{}

\bibitem[\protect\citeauthoryear{Ahmed and Abulaish}{2013}]{ahmed2013generic}
Ahmed, F., and Abulaish, M.
\newblock 2013.
\newblock A generic statistical approach for spam detection in online social
  networks.
\newblock {\em Computer Comms.} 36(10-11):1120--1129.

\bibitem[\protect\citeauthoryear{Al-khateeb and Agarwal}{2019}]{al2019deviance}
Al-khateeb, S., and Agarwal, N.
\newblock 2019.
\newblock {\em Deviance in Social Media and Social Cyber Forensics: Uncovering
  Hidden Relations Using Open Source Information (OSINF)}.
\newblock Springer.

\bibitem[\protect\citeauthoryear{Barrett}{2019}]{barrett2019}
Barrett, P.~M.
\newblock 2019.
\newblock {Disinformation and the 2020 Election: How the social media industry
  should prepare}.
\newblock White paper, Center for Business and Human Rights, New York
  University.

\bibitem[\protect\citeauthoryear{Bessi and Ferrara}{2016}]{bessi2016social}
Bessi, A., and Ferrara, E.
\newblock 2016.
\newblock {Social bots distort the 2016 U.S. Presidential election online
  discussion}.
\newblock {\em First Monday} 21(11).

\bibitem[\protect\citeauthoryear{Blondel \bgroup et al\mbox.\egroup
  }{2008}]{blondel2008fast}
Blondel, V.~D.; Guillaume, J.-L.; Lambiotte, R.; and Lefebvre, E.
\newblock 2008.
\newblock Fast unfolding of communities in large networks.
\newblock {\em Journal of statistical mechanics: theory and experiment}
  2008(10):P10008.

\bibitem[\protect\citeauthoryear{Bovet and Makse}{2019}]{Bovet2019}
Bovet, A., and Makse, H.~A.
\newblock 2019.
\newblock {Influence of fake news in Twitter during the 2016 US presidential
  election}.
\newblock {\em Nature Comms.} 10(1):7.

\bibitem[\protect\citeauthoryear{Cao \bgroup et al\mbox.\egroup
  }{2014}]{cao2014uncovering}
Cao, Q.; Yang, X.; Yu, J.; and Palow, C.
\newblock 2014.
\newblock Uncovering large groups of active malicious accounts in online social
  networks.
\newblock In {\em Proc. of the 2014 ACM SIGSAC Conf. on Computer and Commns.
  Security},  477--488.

\bibitem[\protect\citeauthoryear{Chavoshi, Hamooni, and
  Mueen}{2016}]{chavoshi2016debot}
Chavoshi, N.; Hamooni, H.; and Mueen, A.
\newblock 2016.
\newblock Debot: Twitter bot detection via warped correlation.
\newblock In {\em Proc. Intl. Conf. on Data Mining (ICDM)},  817--822.

\bibitem[\protect\citeauthoryear{Chen and
  Subramanian}{2018}]{chen2018unsupervised}
Chen, Z., and Subramanian, D.
\newblock 2018.
\newblock An unsupervised approach to detect spam campaigns that use botnets on
  twitter.
\newblock {\em arXiv preprint arXiv:1804.05232}.

\bibitem[\protect\citeauthoryear{Ciampaglia \bgroup et al\mbox.\egroup
  }{2018}]{Ciampaglia2018}
Ciampaglia, G.~L.; Mantzarlis, A.; Maus, G.; and Menczer, F.
\newblock 2018.
\newblock {Research challenges of digital misinformation: Toward a trustworthy
  web}.
\newblock {\em AI Magazine} 39(1):65--74.

\bibitem[\protect\citeauthoryear{Cresci \bgroup et al\mbox.\egroup
  }{2016}]{cresci2016dna}
Cresci, S.; Di~Pietro, R.; Petrocchi, M.; Spognardi, A.; and Tesconi, M.
\newblock 2016.
\newblock Dna-inspired online behavioral modeling and its application to
  spambot detection.
\newblock {\em IEEE Intelligent Systems} 31(5):58--64.

\bibitem[\protect\citeauthoryear{Cresci \bgroup et al\mbox.\egroup
  }{2017}]{cresci2017paradigm}
Cresci, S.; Di~Pietro, R.; Petrocchi, M.; Spognardi, A.; and Tesconi, M.
\newblock 2017.
\newblock The paradigm-shift of social spambots: Evidence, theories, and tools
  for the arms race.
\newblock In {\em Proc. of the 26th Intl. Conf. on World Wide Web companion},
  963--972.

\bibitem[\protect\citeauthoryear{Davis \bgroup et al\mbox.\egroup
  }{2016}]{davis2016botornot}
Davis, C.~A.; Varol, O.; Ferrara, E.; Flammini, A.; and Menczer, F.
\newblock 2016.
\newblock Botornot: A system to evaluate social bots.
\newblock In {\em Proc. 25th Intl. Conf. Companion on World Wide Web},
  273--274.

\bibitem[\protect\citeauthoryear{Deb \bgroup et al\mbox.\egroup
  }{2019}]{deb2019perils}
Deb, A.; Luceri, L.; Badaway, A.; and Ferrara, E.
\newblock 2019.
\newblock {Perils and Challenges of Social Media and Election Manipulation
  Analysis: The 2018 US Midterms}.
\newblock In {\em Companion Proc. of The World Wide Web Conf.},  237--247.

\bibitem[\protect\citeauthoryear{Echeverria and
  Zhou}{2017}]{echeverria2017discoveryB}
Echeverria, J., and Zhou, S.
\newblock 2017.
\newblock Discovery, retrieval, and analysis of the'star wars' botnet in
  twitter.
\newblock In {\em Proc. of the 2017 IEEE/ACM Intl. Conf. on Adv. in Social
  Networks Anal. and Mining 2017},  1--8.

\bibitem[\protect\citeauthoryear{Echeverria, Besel, and
  Zhou}{2017}]{echeverria2017discoveryA}
Echeverria, J.; Besel, C.; and Zhou, S.
\newblock 2017.
\newblock Discovery of the twitter bursty botnet.
\newblock {\em Data Science for Cyber-Security}.

\bibitem[\protect\citeauthoryear{Ferrara \bgroup et al\mbox.\egroup
  }{2016}]{ferrara2016rise}
Ferrara, E.; Varol, O.; Davis, C.; Menczer, F.; and Flammini, A.
\newblock 2016.
\newblock {The rise of social bots}.
\newblock {\em Comm. of the ACM} 59(7):96--104.

\bibitem[\protect\citeauthoryear{Ferrara}{2017}]{ferrara2017disinformation}
Ferrara, E.
\newblock 2017.
\newblock {Disinformation and social bot operations in the run up to the 2017
  French presidential election}.
\newblock {\em First Monday} 22(8).

\bibitem[\protect\citeauthoryear{Fortunato}{2010}]{fortunato2010community}
Fortunato, S.
\newblock 2010.
\newblock Community detection in graphs.
\newblock {\em Physics reports} 486(3-5):75--174.

\bibitem[\protect\citeauthoryear{Giglietto \bgroup et al\mbox.\egroup
  }{2020}]{giglietto2020}
Giglietto, F.; Righetti, N.; Rossi, L.; and Marino, G.
\newblock 2020.
\newblock It takes a village to manipulate the media: coordinated link sharing
  behavior during 2018 and 2019 italian elections.
\newblock {\em Information, Communication \& Society} 23(6):867--891.

\bibitem[\protect\citeauthoryear{Grimme, Assenmacher, and
  Adam}{2018}]{grimme2018changing}
Grimme, C.; Assenmacher, D.; and Adam, L.
\newblock 2018.
\newblock Changing perspectives: Is it sufficient to detect social bots?
\newblock In {\em Proc. Intl. Conf. on Social Computing and Social Media},
  445--461.

\bibitem[\protect\citeauthoryear{Grinberg \bgroup et al\mbox.\egroup
  }{2019}]{Grinberg2019}
Grinberg, N.; Joseph, K.; Friedland, L.; Swire-Thompson, B.; and Lazer, D.
\newblock 2019.
\newblock {Fake news on Twitter during the 2016 U.S. presidential election.}
\newblock {\em Science} 363(6425):374--378.

\bibitem[\protect\citeauthoryear{Hills}{2019}]{HillsProliferation18}
Hills, T.~T.
\newblock 2019.
\newblock The dark side of information proliferation.
\newblock {\em Perspectives on Psychological Science} 14(3):323--330.

\bibitem[\protect\citeauthoryear{Hui \bgroup et al\mbox.\egroup
  }{2019}]{huibotslayer}
Hui, P.-M.; Yang, K.-C.; Torres-Lugo, C.; Monroe, Z.; McCarty, M.; Serrette,
  B.; Pentchev, V.; and Menczer, F.
\newblock 2019.
\newblock Botslayer: real-time detection of bot amplification on twitter.
\newblock {\em Journal of Open Source Software} 4(42):1706.

\bibitem[\protect\citeauthoryear{Jowett and O'Donnell}{2018}]{Jowett2018}
Jowett, G., and O'Donnell, V.
\newblock 2018.
\newblock {\em {Propaganda {\&} persuasion}}.
\newblock SAGE Publications, seventh edition.

\bibitem[\protect\citeauthoryear{Keller \bgroup et al\mbox.\egroup
  }{2017}]{keller2017manipulate}
Keller, F.~B.; Schoch, D.; Stier, S.; and Yang, J.
\newblock 2017.
\newblock How to manipulate social media: Analyzing political astroturfing
  using ground truth data from south korea.
\newblock In {\em Eleventh Intl. AAAI Conf. on Web and Social Media}.

\bibitem[\protect\citeauthoryear{Keller \bgroup et al\mbox.\egroup
  }{2019}]{keller2019political}
Keller, F.~B.; Schoch, D.; Stier, S.; and Yang, J.
\newblock 2019.
\newblock Political astroturfing on twitter: How to coordinate a disinformation
  campaign.
\newblock {\em Political Communication}  1--25.

\bibitem[\protect\citeauthoryear{Keogh and
  Ratanamahatana}{2005}]{keogh2005exact}
Keogh, E., and Ratanamahatana, C.~A.
\newblock 2005.
\newblock Exact indexing of dynamic time warping.
\newblock {\em Knowledge and information systems} 7(3):358--386.

\bibitem[\protect\citeauthoryear{Lazer \bgroup et al\mbox.\egroup
  }{2018}]{Lazer-fake-news-2018}
Lazer, D.; Baum, M.; Benkler, Y.; Berinsky, A.; Greenhill, K.; Menczer, F.;
  Metzger, M.; Nyhan, B.; Pennycook, G.; Rothschild, D.; Schudson, M.; Sloman,
  S.; Sunstein, C.; Thorson, E.; Watts, D.; and Zittrain, J.
\newblock 2018.
\newblock The science of fake news.
\newblock {\em Science} 359(6380):1094--1096.

\bibitem[\protect\citeauthoryear{Lee, Eoff, and Caverlee}{2011}]{lee2011seven}
Lee, K.; Eoff, B.~D.; and Caverlee, J.
\newblock 2011.
\newblock Seven months with the devils: A long-term study of content polluters
  on twitter.
\newblock In {\em Fifth International AAAI Conf. on Weblogs and Social Media}.

\bibitem[\protect\citeauthoryear{Mariconti \bgroup et al\mbox.\egroup
  }{2017}]{Mariconti2017}
Mariconti, E.; Onaolapo, J.; Ahmad, S.~S.; Nikiforou, N.; Egele, M.;
  Nikiforakis, N.; and Stringhini, G.
\newblock 2017.
\newblock {What's in a name? Understanding profile name reuse on Twitter}.
\newblock In {\em Proc. 26th Intl. World Wide Web Conf.},  1161--1170.

\bibitem[\protect\citeauthoryear{Miller \bgroup et al\mbox.\egroup
  }{2014}]{miller2014twitter}
Miller, Z.; Dickinson, B.; Deitrick, W.; Hu, W.; and Wang, A.~H.
\newblock 2014.
\newblock Twitter spammer detection using data stream clustering.
\newblock {\em Information Sciences} 260:64--73.

\bibitem[\protect\citeauthoryear{Mirtaheri \bgroup et al\mbox.\egroup
  }{2019}]{mirtaheri2019identifying}
Mirtaheri, M.; Abu-El-Haija, S.; Morstatter, F.; Steeg, G.~V.; and Galstyan, A.
\newblock 2019.
\newblock Identifying and analyzing cryptocurrency manipulations in social
  media.
\newblock {\em arXiv preprint arXiv:1902.03110}.

\bibitem[\protect\citeauthoryear{Pacheco, Flammini, and
  Menczer}{2020}]{pacheco2020whitehelmets}
Pacheco, D.; Flammini, A.; and Menczer, F.
\newblock 2020.
\newblock Unveiling coordinated groups behind white helmets disinformation.
\newblock In {\em CyberSafety 2020: The 5th Workshop on Computational Methods
  in Online Misbehavior. Companion Proc. Web Conf. (WWW)}.

\bibitem[\protect\citeauthoryear{Pennycook \bgroup et al\mbox.\egroup
  }{2019}]{pennycook2019understanding}
Pennycook, G.; Epstein, Z.; Mosleh, M.; Arechar, A.~A.; Eckles, D.; and Rand,
  D.
\newblock 2019.
\newblock Understanding and reducing the spread of misinformation online.
\newblock {\em PsyArXiv preprint: 3n9u8}.

\bibitem[\protect\citeauthoryear{Sayyadiharikandeh \bgroup et al\mbox.\egroup
  }{2020}]{botometerv4-2020}
Sayyadiharikandeh, M.; Varol, O.; Yang, K.-C.; Flammini, A.; and Menczer, F.
\newblock 2020.
\newblock Detection of novel social bots by ensembles of specialized
  classifiers.
\newblock In {\em Proc. 29th ACM International Conf. on Information \&
  Knowledge Management (CIKM)},  2725--2732.

\bibitem[\protect\citeauthoryear{Shao \bgroup et al\mbox.\egroup
  }{2018}]{Shao2018}
Shao, C.; Ciampaglia, G.~L.; Varol, O.; Yang, K.~C.; Flammini, A.; and Menczer,
  F.
\newblock 2018.
\newblock {The spread of low-credibility content by social bots}.
\newblock {\em Nature Comms.} 9(1):4787.

\bibitem[\protect\citeauthoryear{Stella, Ferrara, and
  De~Domenico}{2018}]{stella2018bots}
Stella, M.; Ferrara, E.; and De~Domenico, M.
\newblock 2018.
\newblock Bots increase exposure to negative and inflammatory content in online
  social systems.
\newblock {\em PNAS} 115(49):12435--12440.

\bibitem[\protect\citeauthoryear{Varol \bgroup et al\mbox.\egroup
  }{2017}]{varol2017online}
Varol, O.; Ferrara, E.; Davis, C.~A.; Menczer, F.; and Flammini, A.
\newblock 2017.
\newblock Online human-bot interactions: Detection, estimation, and
  characterization.
\newblock In {\em Proc. 11th Intl. AAAI Conf. on Web and Social Media}.

\bibitem[\protect\citeauthoryear{Vosoughi, Roy, and
  Aral}{2018}]{vosoughi2018spread}
Vosoughi, S.; Roy, D.; and Aral, S.
\newblock 2018.
\newblock The spread of true and false news online.
\newblock {\em Science} 359(6380):1146--1151.

\bibitem[\protect\citeauthoryear{Weng \bgroup et al\mbox.\egroup
  }{2012}]{Weng2012}
Weng, L.; Flammini, A.; Vespignani, A.; and Menczer, F.
\newblock 2012.
\newblock {Competition among memes in a world with limited attention}.
\newblock {\em Scientific Reports} 2(1):335.

\bibitem[\protect\citeauthoryear{Wilson and Starbird}{2020}]{wilson2020cross}
Wilson, T., and Starbird, K.
\newblock 2020.
\newblock Cross-platform disinformation campaigns: lessons learned and next
  steps.
\newblock {\em Harvard Kennedy School Misinformation Review} 1(1).

\bibitem[\protect\citeauthoryear{Yang \bgroup et al\mbox.\egroup
  }{2019}]{Yang2019botometer}
Yang, K.-C.; Varol, O.; Davis, C.~A.; Ferrara, E.; Flammini, A.; and Menczer,
  F.
\newblock 2019.
\newblock Arming the public with artificial intelligence to counter social
  bots.
\newblock {\em Hum. Beh. and Emerging Techn.} 1(1):48--61.

\bibitem[\protect\citeauthoryear{Yang \bgroup et al\mbox.\egroup
  }{2020}]{Yang2020botometer-lite}
Yang, K.-C.; Varol, O.; Hui, P.-M.; and Menczer, F.
\newblock 2020.
\newblock Scalable and generalizable social bot detection through data
  selection.
\newblock In {\em Proc. 34th AAAI Conf. on Artificial Intelligence (AAAI)}.

\bibitem[\protect\citeauthoryear{Yang, Hui, and Menczer}{2019}]{yang2019bot}
Yang, K.-C.; Hui, P.-M.; and Menczer, F.
\newblock 2019.
\newblock Bot electioneering volume: Visualizing social bot activity during
  elections.
\newblock In {\em Companion Proc. of The 2019 World Wide Web Conf.},  214--217.

\end{thebibliography}

\end{document}